\documentclass[aps,prl,10pt,twocolumn,preprintnumbers,superscriptaddress,nofootinbib,letterpaper,longbibliography]{revtex4-2}

\usepackage{amsmath, amssymb}
\usepackage{graphicx}
\usepackage[caption=false]{subfig}
\usepackage{hyperref}
\hypersetup{colorlinks=true, citecolor=blue, linkcolor=blue, urlcolor=blue}
\usepackage[capitalize]{cleveref}
\newcommand{\orcid}[1]{\href{https://orcid.org/#1}{\includegraphics[width=10pt]{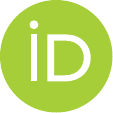}}}

\begin{document}

\title{Towards Powerful Probes of Neutrino Self-Interactions in Supernovae}

\author{Po-Wen Chang \orcid{0000-0003-1134-0652}}           
\email{chang.1750@osu.edu} 
\affiliation{Center for Cosmology and AstroParticle Physics (CCAPP), Ohio State University, Columbus, Ohio 43210}
\affiliation{Department of Physics, Ohio State University, Columbus, Ohio 43210}

\author{Ivan Esteban \orcid{0000-0001-5265-2404}}           
\email{esteban.6@osu.edu}
\affiliation{Center for Cosmology and AstroParticle Physics (CCAPP), Ohio State University, Columbus, Ohio 43210}
\affiliation{Department of Physics, Ohio State University, Columbus, Ohio 43210}

\author{John F. Beacom \orcid{0000-0002-0005-2631}}
\email{beacom.7@osu.edu}
\affiliation{Center for Cosmology and AstroParticle Physics (CCAPP), Ohio State University, Columbus, Ohio 43210}
\affiliation{Department of Physics, Ohio State University, Columbus, Ohio 43210}
\affiliation{Department of Astronomy, Ohio State University, Columbus, Ohio 43210}

\author{Todd A. Thompson \orcid{0000-0003-2377-9574}\,}
\email{thompson.1847@osu.edu}
\affiliation{Center for Cosmology and AstroParticle Physics (CCAPP), Ohio State University, Columbus, Ohio 43210}
\affiliation{Department of Astronomy, Ohio State University, Columbus, Ohio 43210}
\affiliation{Department of Physics, Ohio State University, Columbus, Ohio 43210}

\author{Christopher M. Hirata \orcid{0000-0002-2951-4932}}
\email{hirata.10@osu.edu}
\affiliation{Center for Cosmology and AstroParticle Physics (CCAPP), Ohio State University, Columbus, Ohio 43210}
\affiliation{Department of Physics, Ohio State University, Columbus, Ohio 43210}
\affiliation{Department of Astronomy, Ohio State University, Columbus, Ohio 43210}


\date{\today}

\begin{abstract}

Neutrinos remain mysterious. As an example, enhanced self-interactions ($\nu$SI), which would have broad implications, are allowed.  At the high neutrino densities within core-collapse supernovae, $\nu$SI should be important,  but robust observables have been lacking. We show that $\nu$SI make neutrinos form a tightly coupled fluid that expands under relativistic hydrodynamics.  The outflow becomes either a burst or a steady-state wind; which occurs here is uncertain. Though the diffusive environment where neutrinos are produced may make a wind more likely, further work is needed to determine when each case is realized.  In the burst-outflow case, $\nu$SI increase the duration of the neutrino signal, and even a simple analysis of SN 1987A data has powerful sensitivity.  For the wind-outflow case, we outline several promising ideas that may lead to new observables.  Combined, these results are important steps towards solving the 35-year-old puzzle of how $\nu$SI affect supernovae.

\end{abstract}

\maketitle


\mbox{The weakness of neutrinos makes them powerful~\cite{Raffelt:1996book, Giunti:2007ry, Lesgourgues:2013sjj}.} Because of their near-lack of particle properties, they are a sensitive probe of new physics.  Because of their high abundance, they are a sensitive probe of cosmology.  And because of their penetrating power, they are a sensitive probe of dense sources in astrophysics. Increasingly, progress in one area connects to the others, especially for testing novel-physics scenarios.

An important example is neutrinos with enhanced self-interactions ($\nu$SI, also known as secret interactions as they affect only neutrinos)~\cite{Gelmini:1980re, Georgi:1981pg, Kolb:1987qy, Chacko:2003dt, Davoudiasl:2005ks, Wang:2006jy, Gabriel:2006ns, Bai:2015ztj, Berryman:2018ogk, Blinov:2019gcj, Chacko:2020zze, Dev:2021axj, Kolb:1981mc, Fuller:1988ega, Aharonov:1988ee, Choi:1987sd, Grifols:1988fg, Konoplich:1988mj, Berezhiani:1989za, Choi:1989hi, Akita:2022etk, Blennow:2008er, Dighe:2017sur, Das:2017iuj, Shalgar:2019rqe, Brdar:2020nbj, Hooper:2007jr, Ng:2014pca, Ioka:2014kca, Farzan:2014gza, Murase:2019xqi, Bustamante:2020mep, Esteban:2021tub, Hannestad:2004qu, Hannestad:2005ex, Bell:2005dr, Barenboim:2019tux, DeGouvea:2019wpf}, reviewed in Ref.~\cite{Berryman:2022hds}.  Laboratory probes allow strong $\nu$SI --- orders of magnitude stronger than weak interactions --- and these have been invoked to explain various anomalies~\cite{Araki:2015mya, Liu:2021kug, Lancaster:2017ksf, Kreisch:2019yzn, RoyChoudhury:2020dmd, Escudero:2019gvw, Dentler:2019dhz, Dasgupta:2021ies}.  Cosmological probes also allow strong $\nu$SI, such that early universe physics could be substantially changed. Future astrophysical probes, for example those based on high-energy neutrino propagation through the cosmic neutrino background, will be sensitive to $\nu$SI~\cite{Shoemaker:2015qul, Murase:2019xqi, Esteban:2021tub}.

In principle, core-collapse supernovae should be a powerful probe of $\nu$SI, as the high neutrino densities ($\gtrsim 10^{36} \, \mathrm{cm}^{-3}$) would cause frequent $\nu\text{--}\nu$ scattering (even Standard Model self scattering is non-negligible in supernovae~\cite{Flowers:1976kb, Buras:2002wt, Kotake:2018ypf}). But 35 years after SN 1987A~\cite{Kamiokande-II:1987idp, PhysRevD.38.448, Bionta:1987qt, IMB:1988suc, Arnett:1989tnf}, we still lack robust observables.  The claim by Manohar~\cite{Manohar:1987ec} that $\nu$SI would hinder neutrino escape from the proto-neutron star (PNS) was rebutted by Dicus \emph{et al.}~\cite{Dicus:1988jh}; we discuss both papers below.  Other constraints are weak, have large uncertainties, or rely on future data~\cite{Kolb:1987qy, Blennow:2008er, Das:2017iuj, Dighe:2017sur, Shalgar:2019rqe}. Nevertheless, it is easy to worry that the effects of $\nu$SI could be large enough to alter our deductions about neutrinos and supernovae. New work is needed.

\begin{figure}[b]
    \centering
    \vspace{-0.35cm}
    \includegraphics[width=0.9\columnwidth]{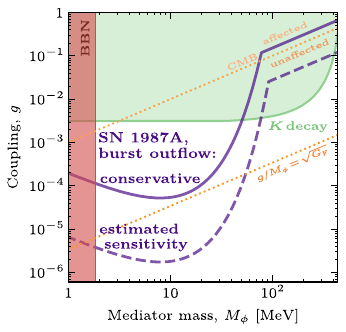}
    \vspace{-0.35cm}
    \caption{Potential constraints on $\nu$SI from SN 1987A (assuming the burst-outflow case), previous limits, and relevant scales~\cite{Arbey:2011nf, Arbey:2018zfh, Berryman:2018ogk, Esteban:2021tub}. $K$-decay bounds apply only to $\nu_e$ and $\nu_\mu$. \emph{Strong $\nu$SI would change the time profile of the SN 1987A neutrino signal}; we show a \textbf{conservative analysis} (30-s duration), and an \textbf{estimated sensitivity} (3-s smearing).}
    \label{fig:bounds}
\end{figure}

In this paper, we re-examine this problem, producing a major first step and a roadmap for the next ones.  We show that for strong $\nu$SI, even self-scattering \emph{outside the PNS} leads to a tightly coupled, expanding neutrino fluid.  There are two possible cases for the outflow --- a burst or a steady-state wind --- and further work is needed to decide when each obtains.  In the burst-outflow case, the {\it observed neutrino signal duration} is a powerful, model-independent probe of $\nu$SI. The neutrino fluid would have a radial extent much greater than the PNS, with individual neutrinos moving in all directions. When decoupling begins, at a time that depends on the $\nu$SI strength, neutrinos would free-stream towards the Earth from the \emph{whole} extended fluid, leading to a longer signal than observed for SN1987A. In the wind-outflow case, decoupling would take place much closer to the PNS.  We will explore this separately, though here we note promising ideas.

\begin{figure*}[htbp]  
    \centering 
    \includegraphics[width=0.9\textwidth]{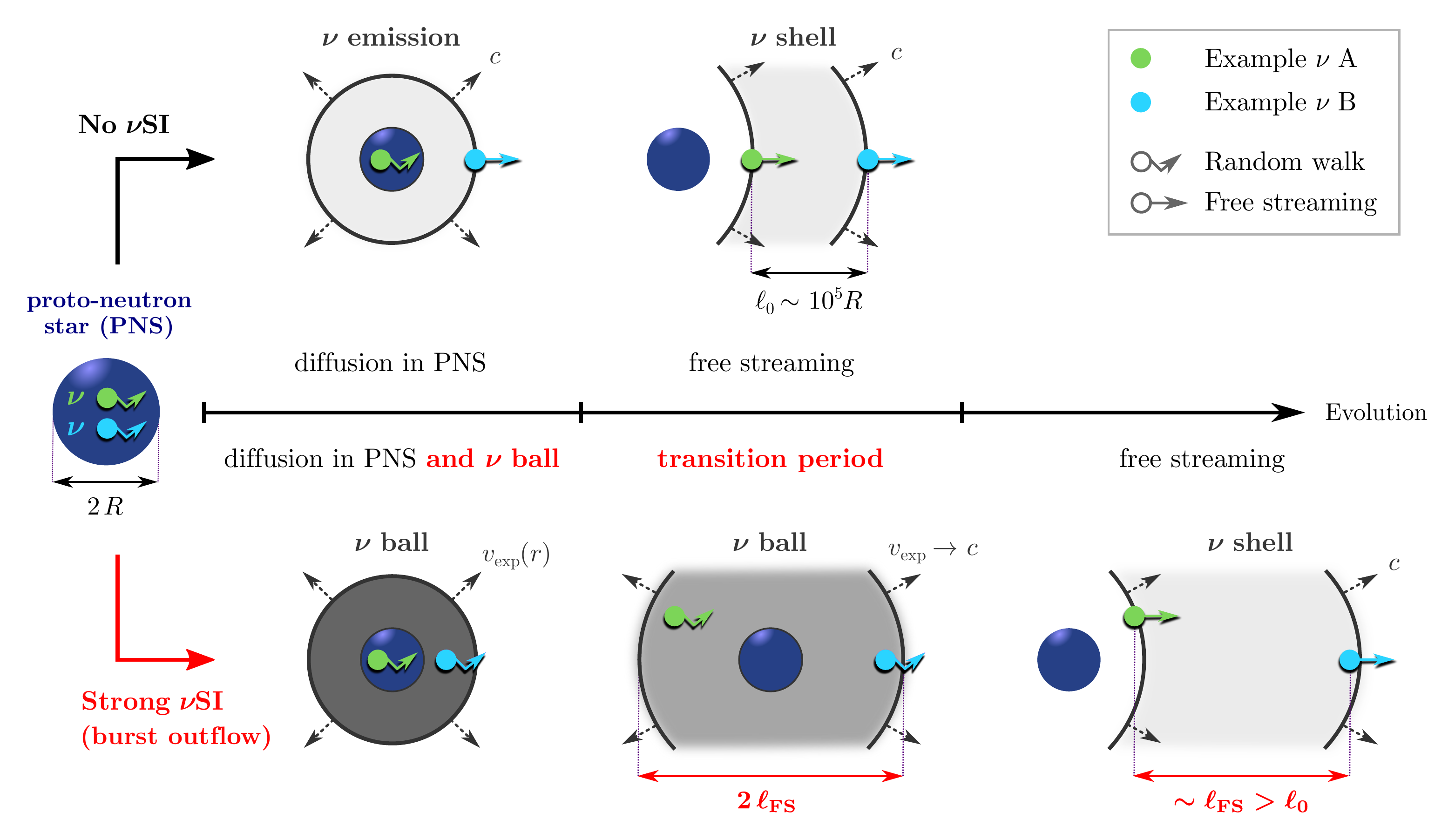}
    \caption{Macroscopic evolution of a neutrino outflow from a supernova (lengths not to scale).  {\bf Without $\boldsymbol{\nu}$SI}, the final width of the neutrino shell is $\ell_0 \sim c \cdot 10 \, \mathrm{s}$, much larger than the PNS and set by neutrino diffusion therein. {\bf With strong $\boldsymbol{\nu}$SI}, neutrinos diffuse in the expanding neutrino ball.  In the burst-outflow case, the size of the ball when neutrinos start decoupling from each other, $\ell_{\rm FS}$, sets the final width of the neutrino shell.  {\it The duration of the observed neutrino signal will thus be significantly extended when $\ell_{\rm FS} > \ell_0$}.
    }
    \label{fig:macro_cartoon}   
\end{figure*}

\Cref{fig:bounds} previews our results for the burst-outflow case, which we focus on in this first paper.  In the following, we review supernova neutrino emission, discuss the impact of $\nu$SI, calculate how they affect the signal duration, contrast this with SN 1987A data, and conclude by outlining future directions. Our approach is simple but conservative, aiming for factor-two precision.  In Supplemental Material (S.M.) that includes Refs.~\cite{Kachelriess:2000qc, Farzan:2002wx, Heurtier:2016otg, Huang:2017egl, IceCube:2020wum, Dighe:1999bi, Dasgupta:2016dbv, Esteban:2020cvm, Bergmann:1999rz, Ge:2018uhz, Burrows:1984zz, Burrows:1986me, Burrows:1990ts, moller_1945, 1975ctf..book.....L, Keil:2002in, Tamborra:2012ac, Goudsmit:1940zza, Burgess:1993xh, Weinberg:1972kfs, 1980Ap&SS..72..447Y, 1965SSRv....4..666P, 1960ratr.book.....C}, we show more detailed calculations and assess the impact of our assumptions.


{\bf\emph{Supernova Neutrino Emission Without $\boldsymbol{\nu}$SI.---}}
For orientation, we describe the basic features of supernova neutrino emission; details are given in S.M. The broad agreement of these predictions with SN 1987A data sets the stage to probe $\nu$SI.  Our estimates are confirmed by supernova simulations that include many important complications~\cite{Iwamoto:1982zp, 2017hsn..book.1575J, Kotake:2012iv, Lentz:2015nxa, Janka:2016fox, Roberts:2016lzn, Burrows:2019zce, Muller:2016izw}.

A supernova begins when electron capture and nuclear photodissociation rob the massive star’s core of pressure support, leading to runaway collapse~\cite{Burbidge:1957vc, Kotake:2005zn, Janka:2012wk, Burrows:2012ew}.
The outcome of the collapse is a compact PNS with a mass ${M \sim 1.5 \, M_\odot}$ and a radius ${R \sim 10 \, \mathrm{km}}$. The collapse leads to a loss of gravitational potential energy of the core $|\Delta E_\mathrm{b}| \sim 3 \times 10^{53} \, \mathrm{ergs}$. Ultimately, almost all of this energy is released in neutrinos.

These neutrinos diffuse through matter until they reach the neutrinosphere, where they decouple and escape. As diffusion suppresses energy flow, their average energy outside the PNS is $\left \langle E_\nu \right \rangle \sim 10 \, \mathrm{MeV}$~\cite{1986rpa..book.....R}. Due to diffusion, the neutrino signal duration is $\sim 10\, \mathrm{s}$~\cite{Arnett:1977xj, 1975PThPh..53..595S, Sato:1975vu, Raffelt:1996book}. Far outside the PNS, this ultimately results in a neutrino shell of thickness ${\ell_0 \simeq c \cdot 10\,{\rm s}}$ that free-streams away at the speed of light.


{\bf\emph{Supernova Neutrino Emission With $\boldsymbol{\nu}$SI.---}}
Due to enhanced $\nu\text{--}\nu$ elastic scattering because of $\nu$SI, \emph{neutrinos do not free-stream after exiting the PNS}. This happens because, as we quantify below, the mean free path is initially tiny, on the $\mu\mathrm{m}$ scale. Neutrinos emitted in all outward directions from each surface element of the PNS promptly scatter with each other. This makes them move in all directions, including inwards, under a random walk (see S.M., where we also discuss how the process conserves momentum). Macroscopically, the coupled neutrino fluid, denoted as the $\nu$ \emph{ball} below, expands as a pressurized gas in vacuum. On the relevant length scales --- much larger than the mean free path --- the behavior of the ball is described by relativistic hydrodynamics. As we detail in S.M., there are two cases to consider.

If, similar to the setup in Dicus \emph{et al.}~\cite{Dicus:1988jh}, we consider the sudden free expansion of a fluid in vacuum, we obtain a burst-like outflow. The ball stays homogeneous, with a near-constant density that decreases as it expands. Any density gradient would rapidly vanish due to the associated pressure difference.  We have verified this with the \texttt{PLUTO} hydrodynamics code~\cite{Mignone_2007}, where we also find that the asymptotic expansion is homologous, i.e., $v_{\rm exp}(r) \propto r$ inside the ball, with $v_{\rm exp} = c$ at the outer boundary. Microscopically, homogeneity is ensured by the random walks mentioned above: any void would be rapidly filled by the randomly moving surrounding neutrinos.

If, on the contrary, given the diffusive nature of the outflow \emph{inside} the PNS, we consider the steady-state case, there is a unique solution, a wind analogous to the well-known relativistic fireball~\cite{Piran:1993jm} (see details in S.M.).  Then the outflow is very different from the burst case, as individual neutrino motions become radial relatively close to the PNS, causing the density outside it to fall as $\sim r^{-2}$. Diffusive systems tend to reach steady-state solutions, but further work is needed to understand the conditions and timescales under which a wind may develop. We outline possible observables below, which will develop in a separate paper. 

\Cref{fig:macro_cartoon} shows how the neutrino fluid evolves without or with $\nu$SI in the burst-outflow case. With $\nu$SI, the neutrino ball expands homogeneously, with a near-constant density that decreases as it expands (bottom left). \emph{The scattering between neutrinos within the ball ends when expansion sufficiently dilutes the density.} We denote the radius of the ball when decoupling begins as $\ell_{\rm FS}$ (bottom center). At this stage, neutrinos go out in all directions (decoupling is almost instantaneous; see below).  The ball then becomes a free-streaming shell with thickness $\sim \ell_{\rm FS}$, from which neutrinos ultimately move radially outward (bottom right). \emph{The critical impact of $\nu$SI on supernova neutrino emission is now clear:} they introduce a new length scale, $\ell_{\rm FS}$, that depends on the $\nu$SI strength  and thus connects the macroscopic behavior of the fluid with the microphysics of $\nu\text{--}\nu$ scattering.  The neutrino signal duration with strong $\nu$SI, $\sim \ell_{\rm FS}/c > \ell_0/c$ ($\ell_0 \sim c \cdot 10 \, \mathrm{s} \sim 10^5 R$), is significantly lengthened. In the wind-outflow case, the neutrino fluid would not be homogeneous nor would neutrinos move in all directions, hence this argument does not apply.

In earlier work, the effects of $\nu$SI on supernova timing were debated, leading to a community consensus that this observable does not provide limits. Manohar~\cite{Manohar:1987ec} claimed that $\nu$SI hinder neutrinos from escaping the PNS, and that the signal duration would be given by the $\nu\text{--}\nu$ diffusion time inside the PNS. In turn, Dicus \emph{et al.}~\cite{Dicus:1988jh} argued that a tightly coupled fluid expands no matter how strong its self-interactions are, hence no limit could be obtained. However, for the burst outflow, the observed duration is set by the size of the neutrino ball at decoupling, which \emph{does depend on the $\nu$SI cross section}, as we compute next. This would lead to powerful new sensitivity.


{\bf\emph{Sensitivity to $\boldsymbol{\nu}$SI Models.---}}
Here we describe our approach (burst-outflow case) to compute $\ell_\mathrm{FS}$, relate it to the $\nu$SI cross section, and constrain $\nu$SI models.

\begin{figure}[t]
    \centering
    \includegraphics[width=0.8\columnwidth]{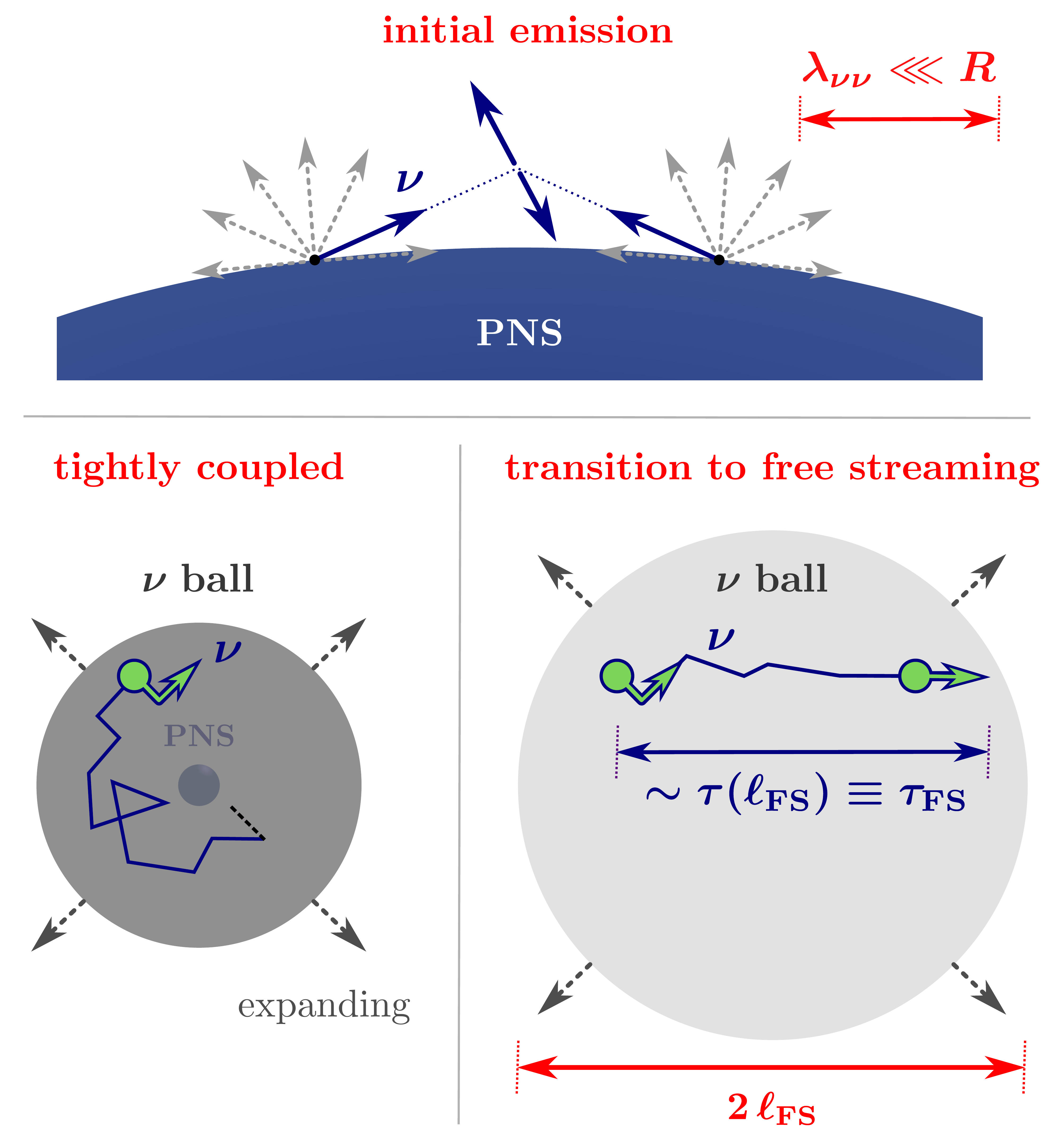}
    \caption{Microscopic evolution of neutrino scattering due to $\nu$SI at different times for the burst-outflow case (lengths not to scale). \emph{Neutrinos move in all directions until the $\nu$SI optical depth becomes small. After then, neutrinos are no longer significantly deflected and the ball becomes a shell.}
    }
    \label{fig:micro_cartoon}
\end{figure}

\Cref{fig:micro_cartoon} illustrates the microphysics. As we discuss above, scattering makes neutrinos move in all directions. Decoupling begins when $\tau$ is small, with $\tau=\tau(\ell)$ the $\nu$SI optical depth, as the number of scatterings a neutrino will undergo when traveling a distance $\ell$ is $\sim \tau^2$.  We denote the optical depth at this stage as ${\tau_\mathrm{FS} \equiv \tau(\ell_\mathrm{FS})}$. For $\tau \lesssim \tau_\mathrm{FS}$, the ball becomes a shell.

The average optical depth for a neutrino traveling a distance $\ell$ is
\begin{equation}
\tau(\ell) = \int \left\langle n_{\nu} \sigma_{\nu\nu} \right\rangle \, \mathrm{d}r \sim \left\langle N_{\nu} \sigma_{\nu\nu} \right\rangle \left( \frac{4\pi}{3} \ell^3 \right)^{-1} \ell\, ,
\label{eq:tau}
\end{equation}
where $n_{\nu} \sim N_\nu / (4\pi \ell^3/3)$ is the neutrino number density, $\sigma_{\nu\nu}$ is the $\nu$SI cross section, and $N_{\nu} \sim |\Delta E_{\rm b}|/\langle E_{\nu} \rangle$ is the number of neutrinos in the ball. We take $N_\nu$ to be the same as without $\nu$SI, as $\nu\text{--}\nu$ scattering conserves the neutrino number except for the largest couplings~\cite{Shalgar:2019rqe} (our results are robust against this; see S.M.). The brackets $\langle...\rangle$ denote the average with respect to the neutrino phase space distributions (see S.M.). Since the number of scatterings ($\sim \tau^2$) decreases as $\ell^{-4}$ as the ball expands, decoupling takes place over a short timescale.

In \cref{eq:tau}, $\sigma_{\nu \nu}$ depends on the parameters of the $\nu$SI model. As a general case of high interest, we consider $\nu$SI among active neutrinos parametrized by the Lagrangian ${\mathcal{L}_{\nu \mathrm{SI}} = - 1/2\, g \bar{\nu } \nu \phi}$ (for UV completions, see Refs.~\cite{Berryman:2018ogk, Blinov:2019gcj, Bai:2015ztj, Dentler:2019dhz, Chacko:2020zze}), where $\phi$ is the mediator with mass $M_\phi$, which for simplicity we take to be a scalar.  We consider Majorana neutrinos, hence the $1/2$ factor. Our results also hold for Dirac neutrinos (see S.M.).
We assume flavor-independent $\nu$SI (see S.M.~for generalizations).

For the mediator mass range we consider, the cross section is s-channel dominated~\cite{Ng:2014pca, Esteban:2021tub},
\begin{equation}
\sigma_{\nu \nu} = \frac{g^4}{16\pi}  \frac{s}{(s-M_\phi^2)^2+M_\phi^2 \Gamma^2} \, , 
\label{eq:csec}
\end{equation}
where $\Gamma = g^2 M_\phi/16\pi$ is the scalar decay width and ${s \equiv 2 E_1 E_2 (1 - \cos \theta_{12})}$, with $E_1$ and $E_2$ the energies of the incoming neutrinos and $\theta_{12}$ their relative angle.

For $s(E_1 E_2,\,\theta_{12}) \sim M_{\phi}^2$ (i.e.,~$M_\phi \sim \langle E_\nu \rangle$) $\nu$SI are resonantly enhanced, leading to large effects. Assuming that neutrinos follow a Maxwell-Boltzmann distribution (our results are insensitive to this, see S.M.), \cref{eq:csec,eq:tau} imply an optical depth in the resonant regime of
\begin{equation}
\tau_{\rm res}(\ell) = \left(\frac{3}{2}\right)^7 \! \frac{|\Delta E_{\rm b}|}{6 \langle E_\nu \rangle} \, \frac{g^2}{M_{\phi}^2} \, \frac{1}{\ell^2} ~ \mathcal{F}\left(\frac{M_{\phi}}{\langle E_\nu \rangle} \right) ~,
\label{eq:taures_analytic}
\end{equation}
with $\mathcal{F}(x) \equiv x^5 K_1(3x) / 3$ and $K_1$ the Bessel function. When neutrino emission begins, ${\tau \sim 4 \times 10^{9} \, (\ell/10\, \mathrm{km})}$ at the edge of our conservative sensitivity in \cref{fig:bounds} for typical neutrino densities $\sim 10^{36}\, \mathrm{cm}^{-3}$. This corresponds to a neutrino mean free path $\ell/\tau \sim \mu\mathrm{m}$, as noted above.

Given the optical depth at decoupling,  $\tau_\mathrm{FS}\equiv\tau(\ell_\mathrm{FS})$, the $\nu$SI strength $g$, and the mediator mass $M_\phi$, \cref{eq:taures_analytic} gives an estimate for the signal duration $\ell_\mathrm{FS}$. Conversely, given $\ell_\mathrm{FS}$ and $\tau_\mathrm{FS}$, we calculate the sensitivity to $g$ in the resonant regime,
\begin{align}     
    & g  \sim ~ 6 \times 10^{-5}  \,\left(\frac{\tau_{\rm FS}}{10}\right)^{1/2} \left(\frac{\ell_{\rm FS}/c}{30 \, \mathrm{s}}\right) \left(\frac{M_\phi}{10 \, \mathrm{MeV}}\right)  \, 
   \label{eq:gres}
   \\
    & \left(\frac{|\Delta E_{\rm b}|}{3 \times 10^{53} \, \mathrm{ergs}}\right)^{-1/2} \! \! \left(\frac{\langle E_\nu \rangle}{10 \, \mathrm{MeV}}\right)^{1/2}
    \left[ \frac{\mathcal{F}(M_\phi/\langle E_\nu \rangle)}{\mathcal{F}(1)} \right]^{-1/2}  \! .   \nonumber
\end{align}
Numerically, the last factor in \cref{eq:gres} stays between $1$ and $10$ as long as $M_\phi$ does not deviate from $\langle E_\nu \rangle$ by more than a factor $\sim 5$.  We take into account this variation as well as the  non-resonant sensitivity (see S.M.).

\begin{figure}[t]
    \centering
    \includegraphics[width=0.9\columnwidth]{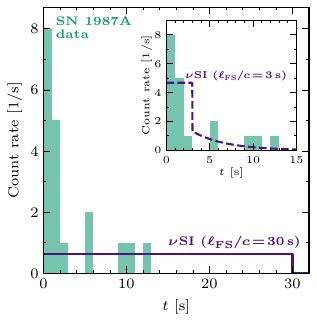}
    \caption{Observed time profile of SN 1987A neutrinos compared to schematic predictions with $\nu$SI for the burst-outflow case. The main figure is for our \textbf{conservative analysis} and the inset for our \textbf{estimated sensitivity}. \emph{$\nu$SI corresponding to the conservative analysis are clearly incompatible with observations}, while those for the estimated sensitivity could be probed with a dedicated analysis.}
    \label{fig:data}
\end{figure}

{\bf\emph{Constraints from SN 1987A.---}}
\Cref{fig:data} shows that, if the burst-outflow case is realized, we can set strong limits on $\nu$SI. For the SN 1987A neutrino data from Kam-II and IMB~\cite{Kamiokande-II:1987idp, PhysRevD.38.448, Bionta:1987qt, IMB:1988suc}, we assume a common start time. Based on the arguments above, the data conservatively exclude $\nu$SI that lead to ${\ell_\mathrm{FS}/c \gtrsim 30 \, \mathrm{s}}$. Even if $\ell_\mathrm{FS}/c$ is smaller than the observed duration $\sim 10 \, \mathrm{s}$, $\nu$SI will still homogenize the neutrino ball, smearing features at times $\lesssim \ell_\mathrm{FS}/c$. A detailed $\nu$SI simulation with a full statistical analysis could probe down to $\ell_\mathrm{FS}/c \sim 3$ s, the smallest timescale at which the data shows clear features.

\Cref{fig:bounds} shows the corresponding sensitivities to $\nu$SI parameters, following the procedure described above.  Because the cross section is largest at the resonance, the sensitivity is best for $M_\phi \sim \langle E_\nu \rangle \sim 10 \, \mathrm{MeV}$.  For our conservative analysis, we assume that decoupling starts when the neutrino optical depth falls below $\tau_\mathrm{FS} = 10$ ($\sim 100$ $\nu\text{--}\nu$ scatterings). For our estimated sensitivity, we take $\tau_\mathrm{FS} = 1$ ($\sim 1$ scattering); then the sensitivity to $g$ in the resonant regime improves by a factor $\sim 30$: a factor 10 from the decrease in $\ell_\mathrm{FS}$, and a factor $\sqrt{10} \sim 3$ from the decrease in $\tau_\mathrm{FS}$. 
In S.M., we display results over a wider mediator mass range and show that decoupling begins for $\tau \lesssim 10$ in our primary region of interest.

If the burst-outflow case is realized, our results are robust. First, we conservatively make minimal assumptions (emission of ${\sim |\Delta E_{\rm b}| / \langle E_\nu \rangle \sim 10^{58}}$ neutrinos with energies ${\sim 10 \, \mathrm{MeV}}$) and focus on the effects of $\nu$SI on $\nu\text{--}\nu$ scattering \emph{far outside the PNS}. Additional effects \emph{inside or near} the PNS (possible extra delays, $2\nu \rightarrow 4 \nu$ processes, neutrino mixing effects, etc.) would either amplify the signal-lengthening signature of $\nu$SI or be subdominant.  Second, as shown in \cref{eq:gres}, the sensitivity to $g$ depends only mildly on the inputs. Third, for even slightly larger $g$ values or earlier times, scattering would be much more frequent (the number of scatterings increases as $\tau^2$, where $\tau \propto g^2/\ell^2$ in the resonant regime and $g^4/\ell^2$ otherwise) and $\nu$SI effects would be enhanced.  Well above our limit, the duration of the signal, scaling as $g/\sqrt{\tau_{\rm FS}}$, would be extreme.  For example, for $M_\phi = 10 \, \mathrm{MeV}$ and $g \sim 10^{-3}$, this would be 10 minutes, leading to an event rate 10 times below Kam-II backgrounds.


{\bf\emph{Conclusions and Future Directions.---}}
Neutrinos are poorly understood and may hold surprises.  An example is $\nu$SI, for which large effects are allowed by laboratory, cosmology, and astrophysics data.  This fact is an opportunity.  It is also a liability, as the effects of $\nu$SI may be biasing our deductions about other physics. As an example, collective mixing effects can be significantly affected by neutrino scattering (reviewed in Ref.~\cite{Capozzi:2022slf}).

In this paper, we re-examine how $\nu$SI affect supernova neutrino emission. We show that the emitted neutrinos form a tightly coupled fluid, with two possible cases for the outflow: burst or wind. Here we focus on the burst case. Although a wind may be more likely, further work is needed to understand when each case obtains.

For the burst-outflow case, we show that the observed duration of a supernova neutrino signal is a robust, powerful signature of $\nu$SI.  Frequent $\nu\text{--}\nu$ scattering outside the PNS leads to a large, tightly coupled, radially expanding ball of neutrinos, internally moving in all directions. This ball decouples with a size depending on the $\nu$SI strength, prolonging and smearing the signal in time.  $\nu$SI causing too long of a duration are strongly excluded, greatly improving upon prior constraints (see~\cref{fig:bounds}). 

Future work may significantly improve sensitivity. Focusing on $\nu$SI effects \emph{far outside} the PNS, the SN 1987A data could be reanalyzed with a detailed $\nu$SI simulation and a full statistical treatment.  For a future galactic supernova, the gains could be much more dramatic, because of the much more precise information on the time profile, flavors, and spectra~\cite{Scholberg:2012id, Mirizzi:2015eza}, which will also solidify the astrophysical model used to test new physics.  Probing the short-timescale features predicted by supernova simulations with high statistics, including the possibility of black hole formation, is especially interesting.  Flavor sensitivity will help probe $\nu$SI strengths in different flavors, complementary to other probes~\cite{Das:2020xke, Brinckmann:2020bcn, Esteban:2021tub}.

For the wind-outflow case, further work and detailed simulations are needed to understand the observable consequences. Relativistic timing effects have been predicted for similar systems~\cite{Piran:2004ba}.  The wind outflow is the \emph{only} steady-state solution to the equations of relativistic hydrodynamics with physical boundary conditions; hence, if it is realized, the entire neutrino fluid \emph{both} outside and inside the PNS would have to relax to it. \emph{Outside} the PNS, this could lead to shocks and other time features that have been observed in numerical explorations of similar systems~\cite{2020MNRAS.493.2834K}.  As a steady-state outflow requires constant energy injection, when the PNS neutrino emission drops~\cite{Li:2020ujl}, a burst outflow could be recovered, leading to potential observables.  \emph{Inside or near} the PNS, the changes could be more dramatic. Differences in the neutrino radial profile between the wind and the no-$\nu$SI cases could affect the supernova. 

In both outflow cases, further observables will likely follow from the physics \emph{inside or near} the PNS. If neutrinos form a tightly coupled fluid, new ways of energy transfer might be possible.  These could affect the temperature and density gradients of matter within the PNS and in the region near the supernova shock.  All of this could be made more complex by changes to neutrino flavor evolution. The sensitivity is potentially exquisite, as at the burst-outflow $\nu$SI limit, the $\nu$SI optical depth inside the PNS is above $\sim 10^{9}$, to be compared to a neutrino-nucleon optical depth of $\sim 10^4$.

The physical conditions in supernovae offer unique opportunities to test both extreme astrophysics and fundamental physics, provided that each is adequately understood.  For 35 years, the impact of $\nu$SI on SN 1987A and future supernovae has been an unsolved puzzle.  A full understanding is needed before the next galactic supernova, so that its data will provide clear new insights.


\bigskip
\begin{acknowledgements}

{\bf\emph{Acknowledgments.---}}
We are grateful for helpful discussions with Vedran Brdar, Francesco Capozzi, Jung-Tsung Li, Shashank Shalgar, Takahiro Sudoh, Bei Zhou, and especially Matheus Hostert, Aneesh Manohar, Thomas Janka, Joachim Kopp, Shirley Li, Georg Raffelt, and Irene Tamborra.   The work of P.-W.C.\ and J.F.B.\ was supported by NSF grant No.\ PHY-2012955 (P.-W.C.\ was also supported by the Studying Abroad Fellowship of Ministry of Education, Taiwan), that of T.A.T.\ by NASA grant No.\ 80NSSC20K0531, and that of C.M.H.\ by NASA award No.\ 15-WFIRST15-0008, Simons Foundation award No.\ 60052667, and the David \& Lucile Packard Foundation. 

\end{acknowledgements}


\bibliography{refs}


\clearpage
\onecolumngrid
\appendix

\ifx \standalonesupplemental\undefined
\setcounter{page}{1}
\setcounter{figure}{0}
\setcounter{table}{0}
\setcounter{equation}{0}
\setcounter{secnumdepth}{2}
\fi

\renewcommand{\thepage}{Supplemental Material -- S\arabic{page}}
\renewcommand{\figurename}{SUPPL. FIG.}
\renewcommand{\tablename}{SUPPL. TABLE}


\clearpage
\newpage

\centerline{\Large {\bf Supplemental Material for}}
\medskip

\centerline{\Large \emph{Towards Powerful Probes of Neutrino Self-Interactions in Supernovae}}
\medskip

\centerline{\large Po-Wen Chang, Ivan Esteban, John F.~Beacom, Todd A.~Thompson, Christopher M.~Hirata}
\bigskip

Here we provide material that is not needed in the main text, but which may help support further developments.  In most sections here, we assume the burst-outflow case, and thus strong constraints based on the timing of the SN 1987A data.  First, we show additional $\nu$SI calculations: \Cref{sec:supplemental_constraints} for a larger range of the parameter space and \Cref{sec:supplemental_flavor} for flavor-dependent scenarios.  Next, we give calculational details for some key results in the main text: \Cref{sec:supplemental_noNuSI} for the basic features of supernova neutrino emission, \Cref{sec:supplemental_tau} for the optical depth calculation and its input-parameter dependence, \Cref{sec:supplemental_isotropization} for scattering establishing random walks and the optical depth below which decoupling begins, and \Cref{sec:supplemental_energy} for how to take into account the energy dependence of the $\nu$SI cross section.  In \Cref{sec:supplemental_Dirac}, we generalize our results to cover Dirac neutrinos.  In the last section, \Cref{sec:supplemental_hydro}, we derive the burst and wind outflow cases from relativistic hydrodynamics, noting some important aspects of the latter case.

When assuming Majorana neutrinos, we use helicity as a proxy for lepton number, i.e., $\nu$ as a shorthand for left-handed and $\bar{\nu}$ as a shorthand for right-handed. This matters because, for resonant scalar production, angular momentum conservation in the center-of-momentum (CM) frame requires the initial neutrinos to have the same helicity. Thus, resonant scattering takes place for $\nu\text{--}\nu$ or $\bar{\nu}\text{--}\bar{\nu}$ interactions.


\section{Constraints over a larger parameter range}
\label{sec:supplemental_constraints}

\begin{figure}[hbtp]
    \centering
    \includegraphics[width=0.88\columnwidth]{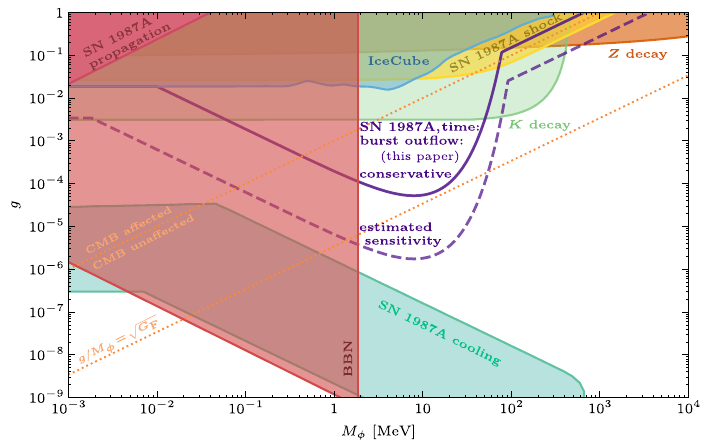}
    \caption{Same as \cref{fig:bounds}, but for a larger range of parameters, now showing important limits from $Z$-boson decay~\cite{Brdar:2020nbj}, SN 1987A cooling~\cite{Kachelriess:2000qc, Farzan:2002wx, Heurtier:2016otg}, and BBN at low $M_\phi$~\cite{Huang:2017egl}.  We also show subdominant limits from present IceCube HESE data~\cite{IceCube:2020wum} (obtained with the code \texttt{nuSIProp} following Ref.~\cite{Esteban:2021tub}), SN 1987A shock revival~\cite{Shalgar:2019rqe}, and SN 1987A neutrino propagation~\cite{Shalgar:2019rqe, Kolb:1987qy}. The CMB line indicates where $\nu$SI delay neutrino free-streaming at the relevant redshifts in the early universe~\cite{Esteban:2021tub}.
    }
    \label{fig:bounds_large}
\end{figure}

Supplemental~\Cref{fig:bounds_large} generalizes \cref{fig:bounds} in the main text to a larger perspective.  We note the attractive possibility of a new analysis of SN 1987A data probing the gap between our timing constraints and prior cooling constraints.

Our estimated sensitivity in $g$ is, for large $M_\phi$, a factor $\sim 10^2$ above the weak scale (given by $g/M_{\phi} = \sqrt{G_{\rm F}}$). Hence, it corresponds to a $\nu$SI cross section $\sigma_{\nu \nu} \sim (10^2)^4 \sigma_{\rm weak} \sim 10^{-35} \, {\rm cm^2}$, with $\sigma_{\rm weak}$ the low-energy weak cross section.

\clearpage
\newpage

\section{Flavor-dependent interactions}
\label{sec:supplemental_flavor}

\begin{figure}[b]
    \centering
    \includegraphics[width=\textwidth]{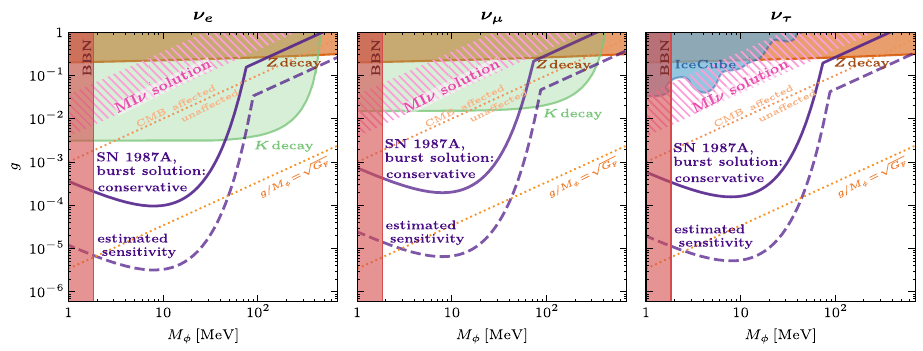}
    \caption{Same as \cref{fig:bounds}, but for $\nu$SI restricted to individual flavors, i.e., assuming $g_{ee}$, $g_{\mu \mu}$ or $g_{\tau \tau}$ to be the only non-zero component.  The ``Moderately Interacting neutrino (MI$\nu$) solution'' (hatched) has been argued to improve the fit to present Cosmic Microwave Background data, but it only remains viable if $\nu$SI are restricted to $\nu_\tau$~\cite{Kreisch:2019yzn}.}
    \label{fig:bounds_flavor}
\end{figure}

Supplemental~\Cref{fig:bounds_flavor} generalizes \cref{fig:bounds} in the main text to flavor-dependent interactions, covering the simple cases where $\nu$SI appears in just one of the three flavor sectors. This is based on new formalism, developed below, where we also discuss flavor structures that could avoid our results.

Though we leave a dedicated study of flavor-dependent $\nu$SI effects for future work, we note some expectations. Because we consider length scales much larger than the oscillation lengths~\cite{Dighe:1999bi,  Dasgupta:2016dbv, Capozzi:2022slf} (including matter effects inside the envelope of the progenitor star, typically larger than the neutrino ball), decoupling naturally happens in the mass-eigenstate basis.  For SN 1987A, only $\bar{\nu}_e$, which are dominantly composed of $\bar{\nu}_1$ and $\bar{\nu}_2$, were detected. Thus, we generically estimate $\ell_\mathrm{FS}$ by setting the optical depths of \emph{both} $\bar{\nu}_1$ and $\bar{\nu}_2$ to $\tau_\mathrm{FS}$, i.e.~(see~\cref{eq:tau}),
\begin{equation}
    \begin{cases}
    \displaystyle{\sum_i \left\langle N_\nu^i \sigma_{\nu\nu}^{1i}\right\rangle \frac{3}{4\pi \ell_\mathrm{FS}^2} = \tau_\mathrm{FS}} \\
    \displaystyle{\sum_i \left\langle N_\nu^i \sigma_{\nu\nu}^{2i}\right\rangle \frac{3}{4\pi \ell_\mathrm{FS}^2} = \tau_\mathrm{FS}}   
    \end{cases} \, .
    \label{eq:tau_FS_flavor}
\end{equation}
This means that strong $\nu$SI could evade our results if only one or none of these mass eigenstates is affected. In this scenario, the constraints from $K$ decay would still apply. Also, the supernova would generate at least two neutrino signals with very different timing and flavor content, a dramatic signature that could be tested with future data.

In \cref{eq:tau_FS_flavor}, $N_\nu^i$ is the number of $\bar{\nu}_i$ in the ball (see the discussion in the preamble of S.M.), and $\sigma_{\nu \nu}^{ji}$ is the cross section for $\bar{\nu}_j \bar{\nu}_i$ interactions. This is dominated by the s-channel contribution~\cite{Esteban:2021tub},
\begin{equation}
\sigma_{\nu \nu}^{ji} = |g_{ij}|^2 \sum_{k, l} |g_{kl}|^2\frac{1}{16\pi}  \frac{s}{(s-M_\phi^2)^2+M_\phi^2 \Gamma^2} \, , 
\end{equation} 
with $g_{ij}$ the elements of the coupling matrix in the mass basis (i.e., $\mathcal{L}_{\nu\mathrm{SI}} = -1/2 \, g_{ij} \bar{ \nu}_i \nu_j$), and $\Gamma = \sum_{k,l} |g_{kl}|^2 \,M_{\phi}/16 \pi$. In the resonant regime, 
\begin{equation}
\sigma_{\nu \nu}^{ji, \, \mathrm{res}} = |g_{ij}|^2 \, \pi \delta(s-M_\phi^2) \, .
\end{equation}
%

Assuming that the supernova emits comparable amounts of neutrinos and antineutrinos of all flavors, $N_\nu^i \sim N_\nu / 6$. Then, our flavor-independent results on $g$ translate into results on $g_\mathrm{eff} \equiv \min\left(\sqrt{\sum_j |g_{1j}|^2}, \, \sqrt{\sum_j |g_{2j}|^2} \right)$ in the resonant regime, and on $g_\mathrm{eff} \equiv \min\left(\sqrt[4]{\sum_j |g_{1j}|^2 \sum_{kl} |g_{kl}|^2}, \, \sqrt[4]{\sum_j |g_{2j}|^2 \sum_{kl} |g_{kl}|^2} \right)$ in the non-resonant regime. The expressions are simpler in the $\nu$SI interaction eigenbasis $\{\nu_a, \nu_b, \nu_c\}$, where our results on $g$ translate into results on 
\begin{equation}
    g_\mathrm{eff} \equiv 
    \begin{cases}
    \min \left( \sqrt{\sum_a g_a^2 |V_{a1}|^2}, \, \sqrt{\sum_a g_a^2 |V_{a2}|^2} \right) & \text{in the resonant regime} \\
    \min \left( \sqrt[4]{\sum_a g_a^2 |V_{a1}|^2 \sum_b g_b^2}, \, \sqrt[4]{\sum_a g_a^2 |V_{a2}|^2 \sum_b g_b^2} \right) & \text{in the non-resonant regime}
    \end{cases} \, ,
    \label{eq:g_eff_flavor}
\end{equation}
where $\{g_a, g_b, g_c\}$ are the eigenvalues of $g_{ij}$, and $V$ is the matrix linking $\nu$SI interaction eigenstates to mass eigenstates as ${\nu_a = \sum_i V_{ai} \nu_i}$. With this expression and the mixing angles from Ref.~\cite{Esteban:2020cvm}, we obtain the results in Suppl.~\cref{fig:bounds_flavor} (for the matrix elements that depend on $\delta_\mathrm{CP}$, we vary this parameter between 0 and $2\pi$ and we conservatively take the largest $g_\mathrm{eff}$).

Finally, we note that flavor-dependent $\nu$SI could introduce an effective $\nu {\emph - }\nu$ potential affecting flavor oscillations. However, for scalar $\nu$SI this is a correction to the neutrino mass $\sim n_\nu (g^2/M_\phi^2) (m_\nu/E_\nu)$~\cite{Bergmann:1999rz, Ge:2018uhz}. The suppression factor for ultrarelativistic neutrinos $m_\nu/E_\nu \sim 10^{-8}$ guarantees that this correction is small.


\clearpage
\newpage

\section{Estimation of supernova neutrino properties}
\label{sec:supplemental_noNuSI}

Here we expand the discussion on the basic features of supernova neutrino emission. We show that simple arguments based on energy conservation and neutrino diffusion through matter predict emission of $|\Delta E_b| \sim 3 \times 10^{53}\,\mathrm{ergs}$ in neutrinos with average energy $\langle E_\nu \rangle \sim 10\,\mathrm{MeV}$ and, in the absence of $\nu$SI, a signal duration $\sim 10\, \mathrm{s}$.

\emph{Neutrino total energy.} Core collapse stops when the density is comparable to nuclear density, ${\rho_\mathrm{n} \sim 3 \times 10^{14} \, \mathrm{g}/\mathrm{cm}^3}$, beyond which the equation of state becomes stiff. The mass of the PNS is similar to that of observed neutron stars, ${M \sim 1.5 \, M_\odot}$, and thus its radius is 
\begin{equation}
    R = \left(\frac{3M}{4 \pi \rho_\mathrm{n}}\right)^{1/3} \sim 10 \, \mathrm{km} \, .
\end{equation}
The loss of gravitational potential energy of the core is
\begin{equation}
    |\Delta E_\mathrm{b}| = \frac{3}{5} \left[ \frac{G_N M^2}{R}\right]_\mathrm{final} -  \frac{3}{5} \left[ \frac{G_N M^2}{ R }\right]_\mathrm{initial} \simeq \frac{3}{5} \left[ \frac{G_N M^2}{R}\right]_\mathrm{final} \sim \frac{3}{5} \frac{G_N \, (1.5 M_\odot)^2 }{10 \, \mathrm{km} } \sim 3 \times 10^{53} \, \mathrm{ergs} \, ,
\end{equation}
where $G_N$ is Newton's constant and we have used $R_\mathrm{final} \ll R_\mathrm{initial}$. Ultimately, almost all of this energy is released in neutrinos, because other particles cannot escape.

\emph{Neutrino average energy.} The collapse raises by ${\sim |\Delta E_{\rm b}|}$ the thermal energy of the nucleons (and electrons), which then produce neutrinos through several processes.  Because neutrinos are in thermal equilibrium with matter due to the large density, the nucleon thermal energy sets the average neutrino energy in the core, 
\begin{equation}
    \left \langle E_\nu^\mathrm{core} \right \rangle \sim \frac{|\Delta E_\mathrm{b}|}{M / m_\mathrm{N}} \sim 100 \, \mathrm{MeV} \, ,
\end{equation}
with $m_\mathrm{N}$ the nucleon mass. Neutrinos then diffuse through matter until they reach the neutrinosphere, where they decouple and escape. Their average energy outside the PNS, $\langle E_\nu \rangle$, is smaller than $\left \langle E_\nu^\mathrm{core} \right \rangle$ because diffusion suppresses energy flow~\cite{1986rpa..book.....R}
\begin{equation}
    \left \langle E_\nu \right \rangle \sim \tau_{\nu \mathrm{N}}^{-1/4} \left \langle E_\nu^\mathrm{core} \right \rangle \, ,
\end{equation}
with $\tau_{\nu \mathrm{N}}\sim R/\lambda_{\nu \mathrm{N}}$ the neutrino optical depth on nucleons and $\lambda_{\nu \mathrm{N}}$ the mean free path. Using the neutrino-nucleon weak interaction cross section, we obtain $\left \langle E_\nu \right \rangle \sim 10 \, \mathrm{MeV}$.

\emph{Neutrino signal duration.} Neutrinos diffuse out of the PNS over timescales $\sim 3R^2 / (c\,\lambda_{\nu {\rm N}})  \sim 1\, \mathrm{s}$~\cite{Arnett:1977xj, 1975PThPh..53..595S, Sato:1975vu, Raffelt:1996book}. The neutrino signal duration is longer by a factor $\sim 10$~\cite{Burrows:1984zz, Burrows:1986me, Burrows:1990ts}, because the stored thermal energy in matter continues to be emitted in neutrinos even after the first neutrinos escape. Far outside the PNS, this ultimately results in a neutrino shell of thickness ${\ell_0 \simeq c \cdot 10\,{\rm s}}$ that free-streams away at the speed of light.


\clearpage
\newpage

\section{Computation of the neutrino optical depth}
\label{sec:supplemental_tau}

Here we compute the average $\nu$SI neutrino optical depth given in \cref{eq:tau}.  Since we aim for factor-two precision, we conservatively consider a neutrino that travels a distance $\ell$ (the radius of the ball) instead of $2\ell$ (the diameter of the ball). We also assume a uniform background density $n_\nu = N_\nu / (4 \pi \ell^3 / 3)$, neglecting $\mathcal{O}(1)$ variations due to the expansion of the ball and its conversion into a shell. The general expression then reads
\begin{equation}
    \tau(\ell) = \frac{1}{6} N_\nu \left(\frac{4 \pi}{3} \ell^3\right)^{-1} \ell \int \mathrm{d}E_1 \, \mathrm{d}E_2 \, \mathrm{d}\Omega_1 \, \mathrm{d}\Omega_2 \, f(E_1, \hat{\Omega}_1) f(E_2, \hat{\Omega}_2) \, (1 - \cos \theta_{12}) \, \sigma_{\nu \nu}(E_1, E_2, \cos \theta_{12}) \, .
    \label{eq:tau_integral}
\end{equation}
The factor $\frac{1}{6}$ takes into account that we consider s-channel scattering and flavor-independent $\nu$SI: as discussed in the preamble of S.M., a neutrino of a given flavor only scatters with another neutrino of the same flavor (and analogously for antineutrinos) and we assume that supernovae emit comparable amounts of neutrinos and antineutrinos of all three flavors. In \cref{eq:tau_integral}, $N_\nu \sim |\Delta E_{\rm b}| / \langle E_\nu \rangle$ is the total number of neutrinos in the ball, $E_1$ and $E_2$ are the energies of the incoming neutrinos, $\Omega_1$ and $\Omega_2$ are the solid angles covered by their directions of motion, $\theta_{12}$ is the angle between their momenta, $f(E, \hat{\Omega})$ is the neutrino energy and angular distribution, $(1- \cos \theta_{12})$ is the M{\o}ller factor~\cite{moller_1945, 1975ctf..book.....L}, and $\sigma_{\nu \nu}$ is the $\nu$SI cross section. 
We assume that neutrinos move isotropically. While this is not strictly true, particularly close to the edges of the expanding neutrino ball, directional effects average when considering interactions among neutrinos from the entire ball. For generality, we take a pinched Maxwell-Boltzmann neutrino energy distribution~\cite{Keil:2002in, Tamborra:2012ac},
\begin{equation}
    f(E, \hat{\Omega}) = \frac{1}{4 \pi}\frac{(\beta + 1)^{\beta + 1}}{\Gamma(\beta + 1)} \frac{E^\beta}{\langle E_\nu \rangle^{\beta +1}} e^{-(\beta+1) \frac{E}{\langle E_\nu \rangle}} \, ,  
    \label{eq:energy_distribution}
\end{equation}
with $\Gamma$ being Euler's function; and $\beta$ a ``pinching'' parameter that controls the width of the distribution, where $\beta=2$ corresponds to a Maxwell-Boltzmann distribution, $\beta=2.3$ to a Fermi Dirac distribution with no chemical potential, and other values give other distributions. Supernova simulation outputs are well described by $\beta \in [2, 4]$~\cite{Tamborra:2012ac, Keil:2002in}. As we show below, our results are robust to changes in $\beta$, so in the main text we assume $\beta=2$.

We consider the s-channel scattering cross section in \cref{eq:csec}. Including other scattering channels would increase both the cross section and the number of target neutrinos (although random walks in all directions may not be guaranteed; see \cref{sec:supplemental_isotropization}). This may lead to improved constraints, particularly for light mediators (see, e.g., Ref.~\cite{Esteban:2021tub}).

For simplicity, we separate our calculations into three different kinematic regimes:
\begin{itemize}
    \item $M_\phi \ll \sqrt{s}$: here,
    \begin{equation}
        \sigma_{\nu \nu} = \frac{g^4}{16 \pi} \frac{1}{s} = \frac{g^4}{32 \pi} \frac{1}{E_1 E_2 (1 - \cos \theta_{12})} \, , 
        \label{eq:csec_nres_SM}
    \end{equation}
    and \cref{eq:tau_integral} gives
    \begin{equation}
        \tau = \frac{3}{128 \pi^2} \left(\frac{1+\beta}{\beta}\right)^2 \, \frac{|\Delta E_{\rm b}|}{6 \langle E_\nu \rangle} \, \frac{g^4}{\langle E_\nu \rangle^2} \, \frac{1}{\ell^2} \, .
         \label{eq:tau_lowM}
    \end{equation}
    As described in the main text, we estimate $\ell_\mathrm{FS}$ as a function of $g$ and $M_\phi$ by taking $\tau(\ell_\mathrm{FS}) \equiv \tau_\mathrm{FS}$. Conversely, given $\ell_\mathrm{FS}$ and $\tau_\mathrm{FS}$, the sensitivity to $g$ is given by
    \begin{equation}
        g \sim 1.9 \times 10^{-2} \,\left(\frac{\tau_{\rm FS}}{10}\right)^{1/4} \left(\frac{\ell_{\rm FS}/c}{30 \, \mathrm{s}}\right)^{1/2} 
        \left(\frac{|\Delta E_{\rm b}|}{3 \times 10^{53} \, \mathrm{ergs}}\right)^{-1/4} \left(\frac{\langle E_\nu \rangle}{10 \, \mathrm{MeV}}\right)^{3/4} \left(\frac{(1+\beta)/\beta}{3/2}\right)^{-1/2} \, .
        \label{eq:sens_lowM}
    \end{equation}
    \item $M_\phi \gg \sqrt{s}$: here,
    \begin{equation}
        \sigma_{\nu \nu} = \frac{g^4}{16 \pi} \frac{s}{M_\phi^4} = \frac{g^4}{8 \pi} \frac{E_1 E_2 (1 - \cos \theta_{12})}{M_\phi^4} \label{eq:csec_nres_LM}
    \end{equation}
    and \cref{eq:tau_integral} gives
    \begin{equation}
        \tau = \frac{1}{8 \pi^2} \, \frac{|\Delta E_{\rm b}|}{6 \langle E_\nu \rangle} \, g^4 \frac{\langle E_\nu \rangle^2}{M_\phi^4} \, \frac{1}{\ell^2} \, .
        \label{eq:tau_largeM}
    \end{equation}
    The sensitivity to $g$ is given by
    \begin{equation}
        g \sim 0.15 \,\left(\frac{\tau_{\rm FS}}{10}\right)^{1/4} \left(\frac{\ell_{\rm FS}/c}{30 \, \mathrm{s}}\right)^{1/2} \left(\frac{M_\phi}{100 \, \mathrm{MeV}}\right)
        \left(\frac{\Delta E_{\rm b}}{3 \times 10^{53} \, \mathrm{ergs}}\right)^{-1/4} \left(\frac{\langle E_\nu \rangle}{10 \, \mathrm{MeV}}\right)^{-1/4} \, .
        \label{eq:sens_largeM}
    \end{equation}
    \item $M_\phi \sim \sqrt{s}$: here,
    \begin{equation}
        \sigma_{\nu \nu} = g^2 \pi \,\delta(s-M_\phi^2) = g^2 \pi \, \delta(2 E_1 E_2 [1 - \cos \theta_{12}] - M_\phi^2) \,,
        \label{eq:csec_res}
    \end{equation}
    and \cref{eq:tau_integral} gives
    \begin{equation}
        \tau = \frac{3}{8}  \left(\frac{(1+\beta)^{1+\beta}}{2^{\beta/2} \Gamma(1+\beta)}\right)^2 \, \frac{|\Delta E_{\rm b}|}{6 \langle E_\nu \rangle} \, \frac{g^2}{M_\phi^2} \, \frac{1}{\ell^2} \, \mathcal{F}_\beta \left( \frac{M_\phi}{\langle E_\nu \rangle} \right) \, ,
        \label{eq:taures_beta}
    \end{equation}
    with 
    \begin{equation}
        \mathcal{F}_\beta (x) \equiv x^{2+2\beta} \int_0^2 \mathrm{d}q \, q^{-\beta} K_0\left(\sqrt{\frac{2}{q}}(1+\beta) x\right) \, , \label{eq:F_beta}
    \end{equation}
    and $K_0$ the modified Bessel function. For a given $\ell_\mathrm{FS}$, the sensitivity to $g$ is given by
    \begin{equation}
    \begin{split}
    g  \sim 5.7 \times 10^{-5}  \, & \left(\frac{\tau_{\rm FS}}{10}\right)^{1/2} \left(\frac{\ell_{\rm FS}/c}{30 \, \mathrm{s}}\right) \left(\frac{M_\phi}{10 \, \mathrm{MeV}}\right)   \left(\frac{|\Delta E_{\rm b}|}{3 \times 10^{53} \, \mathrm{ergs}}\right)^{-1/2} \left(\frac{\langle E_\nu \rangle}{10 \, \mathrm{MeV}}\right)^{1/2}
    \left[ \frac{\mathcal{F}_\beta(M_\phi/\langle E_\nu \rangle)}{\mathcal{F}_2(1)} \right]^{-1/2} \\ & \left( \frac{2^{\beta/2} \Gamma(1+\beta) / (1+\beta)^{1+\beta}}{4/27} \right) \, .
    \label{eq:sens_resonance}
    \end{split} 
    \end{equation}   
    For $\beta=2$, $\mathcal{F}_2(x) = x^5 \,K_1(3x)/3$, and  \cref{eq:taures_beta,eq:sens_resonance} correspond to \cref{eq:taures_analytic,eq:gres} in the main text. 
\end{itemize}
\Cref{eq:sens_largeM,eq:sens_lowM,eq:sens_resonance} give our sensitivity for any mediator mass in \cref{fig:bounds} and Suppl.~\cref{fig:bounds_large}. In the transitions between the different kinematic regimes, we conservatively take the largest optical depth instead of adding them up.

\begin{figure}[hbtp]
    \centering
    \includegraphics[width=\textwidth]{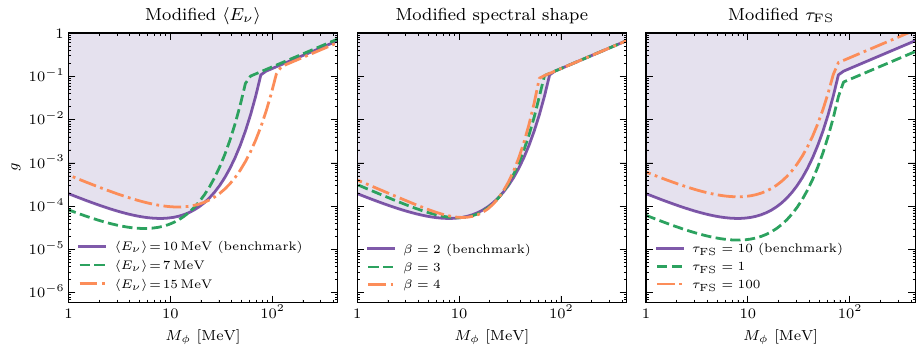}
    \caption{Conservative sensitivity in \cref{fig:bounds} for different assumptions, as labeled. }
    \label{fig:bounds_assumptions}
\end{figure}

Supplemental~\Cref{fig:bounds_assumptions} shows that our results are mostly insensitive to our assumptions. Changing $\langle E_\nu \rangle$ modifies the mediator mass for which resonant scattering is most frequent, $M_\phi \sim \langle E_\nu \rangle$; the cross section for that mediator mass, ${\sigma \propto 1/M_\phi^2 \sim 1/\langle E_\nu \rangle^2}$; as well as the number of neutrinos in the ball, $N_\nu \propto |\Delta E_{\rm b}| / \langle E_\nu \rangle$. These effects do not affect our limit by more than a factor $\sim 2$ even if we allow $\langle E_\nu \rangle$ to change by $\sim 50\%$. Our result is thus insensitive to small variations in the neutrino energy due to the temperature evolution of the supernova, gravitational redshift of neutrinos as they leave the PNS, or hierarchies in the average energies of the different flavors. Changing the pinching parameter $\beta$ modifies the low-energy and high-energy tails of the neutrino energy distribution, which induces minor differences on the resonant production of mediators with masses $M_\phi \neq \langle E_\nu \rangle$.  Finally, even if we are extremely conservative and assume that decoupling begins as soon as $\tau$ drops below $\sim 100$, i.e., setting $\tau_\mathrm{FS} = 100$ instead of $10$, our limit on $g$ only changes by a factor $\sqrt{100/10} \sim 3$ (or $\sqrt[4]{100/10} \sim 1.8$ outside the resonant regime). Changing the total neutrino energy $|\Delta E_{\rm b}|$ is numerically equivalent to changing $\tau_\mathrm{FS}$, as our limit depends on $\tau_\mathrm{FS}/|\Delta E_{\rm b}|$.

\begin{figure}[htbp]
    \centering
    \includegraphics[width=0.60\textwidth]{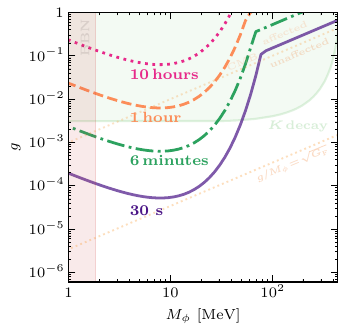}    \caption{Supernova neutrino signal duration with $\nu$SI for the burst-outflow case. Other limits and scales are as in \cref{fig:bounds}.}
    \label{fig:bounds_time}
\end{figure}

Supplemental~\Cref{fig:bounds_time} shows that in the burst-outflow case the supernova neutrino signal duration is extremely large if the coupling is above our conservative limit. We compute the signal duration conservatively assuming that the neutrino ball starts decoupling at optical depth $\tau_\mathrm{FS} = 10$. For a mediator mass $M_\phi \sim 10 \, \mathrm{MeV}$ and a coupling that saturates $K$-decay bounds, the signal duration is ${\sim 1 \, \mathrm{hour}}$; and even greater for some interactions affecting cosmology. 

The results above also demonstrate that our limit for the burst-outflow case is robust against $2 \nu \rightarrow 4 \nu$ processes. These take place only for the largest couplings~\cite{Shalgar:2019rqe} (the yellow region labeled ``SN 1987A shock'' in Suppl.~\cref{fig:bounds_large}), and they reduce the average neutrino energy $\langle E_\nu \rangle$. For $M_\phi \lesssim 70 \, \mathrm{MeV}$, such large couplings imply a supernova neutrino signal duration ${\gtrsim 1 \, \mathrm{hour}}$ (see~Suppl.~\cref{fig:bounds_time}), very strongly excluded no matter what value of $\langle E_\nu \rangle$ is used. For $M_\phi \gtrsim 70 \, \mathrm{MeV}$, our limit scales as ${g \sim \langle E_\nu \rangle^{-1/4}}$, see~\cref{eq:sens_largeM} and Suppl.~\cref{fig:bounds_assumptions}, and it is therefore insensitive to the precise value of $\langle E_\nu \rangle$.


\clearpage
\newpage

\section{Random walks induced by frequent scattering}
\label{sec:supplemental_isotropization}

Here we quantify how, when scattering becomes rare as the neutrino ball dilutes, neutrinos are no longer significantly deflected by $\nu$SI, random motion in all directions is no longer established, and the ball becomes a shell (see~\cref{fig:macro_cartoon,fig:micro_cartoon}).

\begin{figure}[hbtp]
    \centering
    \includegraphics[width=0.68\textwidth]{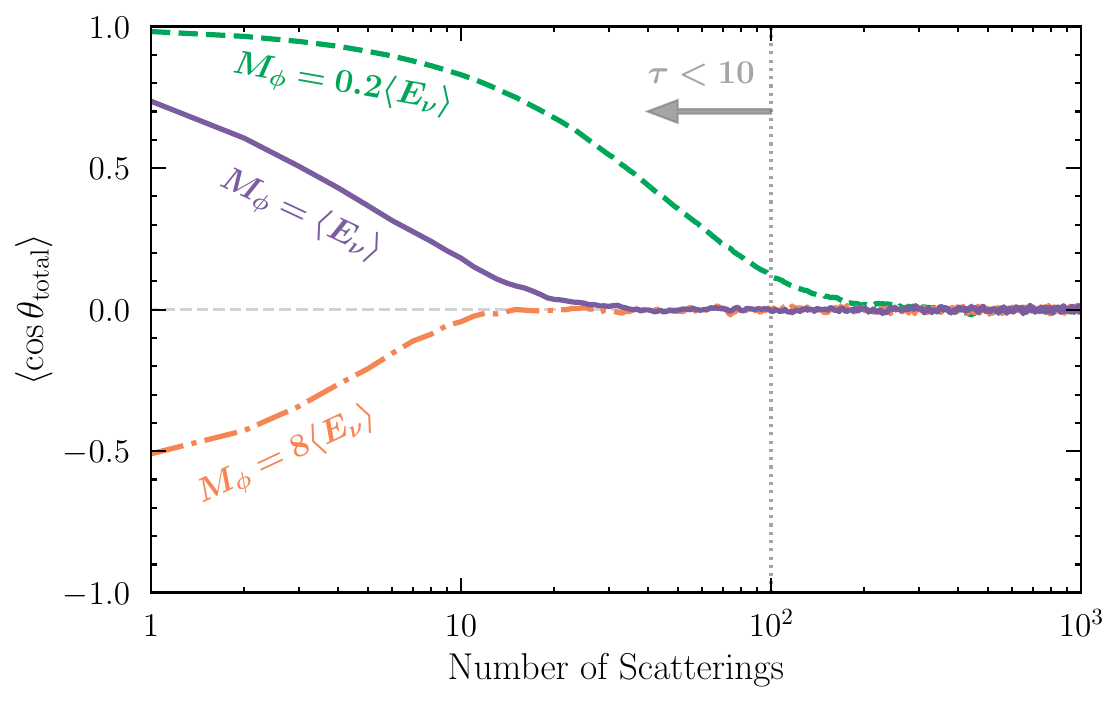}  
    \caption{Average deflection of an initially forward neutrino ensemble after many resonant scatterings with an isotropic background, for different $M_{\phi}/\langle E_{\nu} \rangle$ (see below for non-resonant scattering). When $\langle \cos \theta_\mathrm{total} \rangle\sim 0$, $\nu$SI randomize the directions of motion. As the neutrino ball expands and dilutes, the $x$-axis moves from right to left.}
    \label{fig:def_total}
\end{figure}

\begin{figure}[hbtp]
    \centering
    \includegraphics[width=0.75\textwidth]{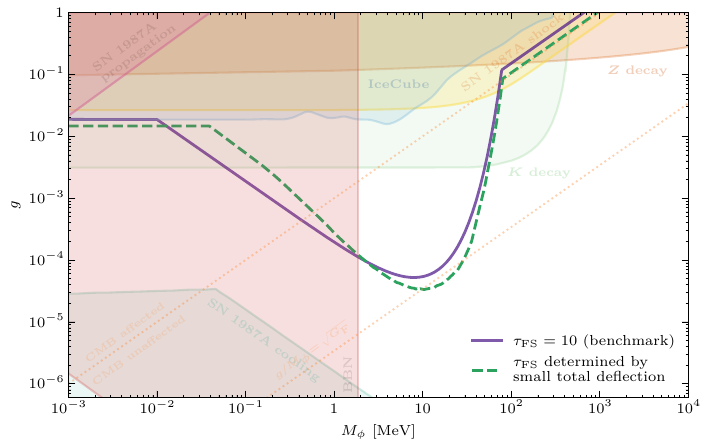}  
    \caption{Conservative sensitivity in Suppl.~\cref{fig:bounds_large} for different choices of $\tau_\mathrm{FS}$ (note the different $y$-axis range). In our primary region of interest, above BBN limits, our choice in the main text is shown to be always conservative.}
    \label{fig:bounds_FS_isotropic}
\end{figure}

Supplemental~\Cref{fig:def_total} shows the number of scatterings below which neutrino deflections are no longer significant.  For mediator masses in our region of interest (see \cref{fig:bounds}), this happens for $\tau \sim 10$. Thus, setting the transition period at $\tau_\mathrm{FS} = 10$ leads to a conservative underestimate of the signal duration. (See Suppl.~\cref{fig:bounds_FS_isotropic} for the impact for a wider mediator mass range.)

In the opposite limit, when the number of scatterings is enormous (as happens in the early emission, see main text and \cref{fig:micro_cartoon}), motion in all directions is quickly established. This is so despite momentum conservation (which introduces a preferred direction) because one can always boost to the frame where net momentum locally vanishes, and there frequent collisions rapidly isotropize directions. The laboratory frame distribution is also quite isotropic because, for the burst solution we consider, the net velocities of the neutrino fluid are small compared to $c$.

Intuitively, the number of scatterings in Suppl.~\cref{fig:def_total} to establish random motion is not very large because for the scalar-mediated s-channel scattering we consider, the distribution of the outgoing neutrino directions is isotropic in the CM frame. Furthermore, the CM boost is
\begin{equation}
    \gamma_\mathrm{CM} = \frac{1}{\sqrt{1 - \beta_\mathrm{CM}^{\,2}}} = \frac{1 + E_1 / E_2}{2 \sqrt{E_1/E_2}} \cdot \left( {\rm sin}\,\frac{\theta_{12}}{2} \right)^{-1} \,, 
    \label{eq:gamma_c}
\end{equation}
which is typically $\mathcal{O}(1)$ as $E_1$ and $E_2$, the energies of the incoming neutrinos, are drawn from similar distributions; and the scattering angle $\theta_{12}$ is not generally small. Thus, the outgoing distribution in the laboratory frame is also quite isotropic. 

For increasing $M_\phi/\langle E_\nu \rangle$, neutrinos move in all directions with less scatterings. There are several ways of understanding this. From energy and momentum conservation, producing a heavier mediator requires a larger scattering angle $\theta_{12}$. This leads to a large momentum transfer, and thus to a large deflection. Alternatively, larger $\theta_{12}$ decreases the CM boost (see~\cref{eq:gamma_c}), making the outgoing neutrinos more isotropic. Finally, one can also check that for resonant scattering the CM boost can be written as ${\gamma_\mathrm{CM}=(E_1+E_2)/\sqrt{s} = (E_1+E_2)/M_\phi}$. Larger $M_\phi$ thus leads to smaller CM boosts and more isotropic scattering.

\begin{figure}[hbtp]
    \centering
    \includegraphics[width=0.62\textwidth]{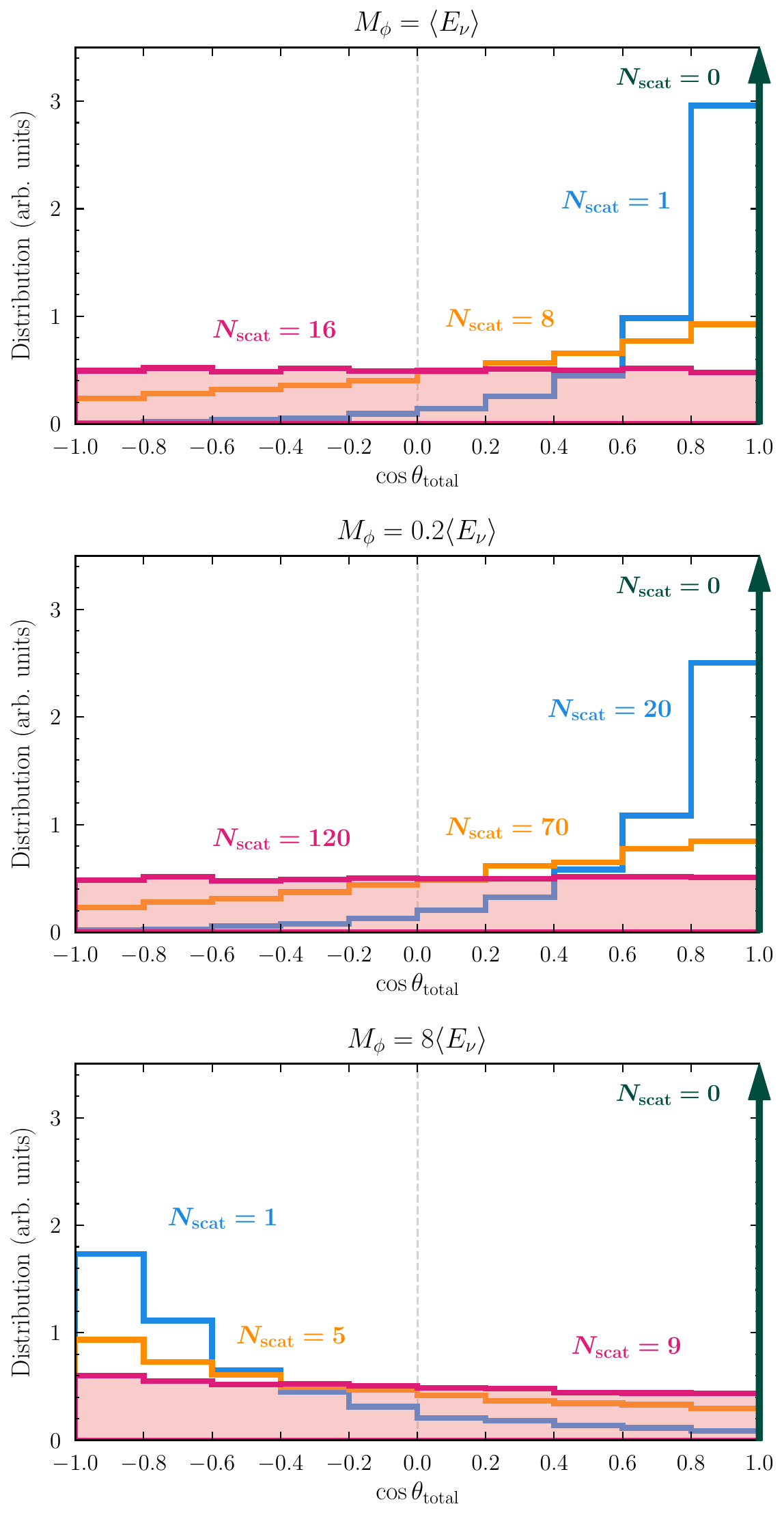}  
    \caption{Distribution of directions of an initially forward neutrino ensemble after many resonant scatterings with an isotropic background, for different $M_\phi/\langle E_{\nu}\rangle$. The total deflection angle, $\theta_{\rm total}$, is measured with respect to the initial direction.}
    \label{fig:def_dist}
\end{figure}

Supplemental \Cref{fig:def_dist} explicitly shows that the distribution of neutrino directions becomes isotropic after scattering. For large $M_\phi$, the momentum transfer in a resonant scattering is large as noted above, leading to a backwards-peaked distribution for the first scatterings (see~top panel in the figure).

\begin{figure}[hbtp]
    \centering
    \includegraphics[width=0.62\textwidth]{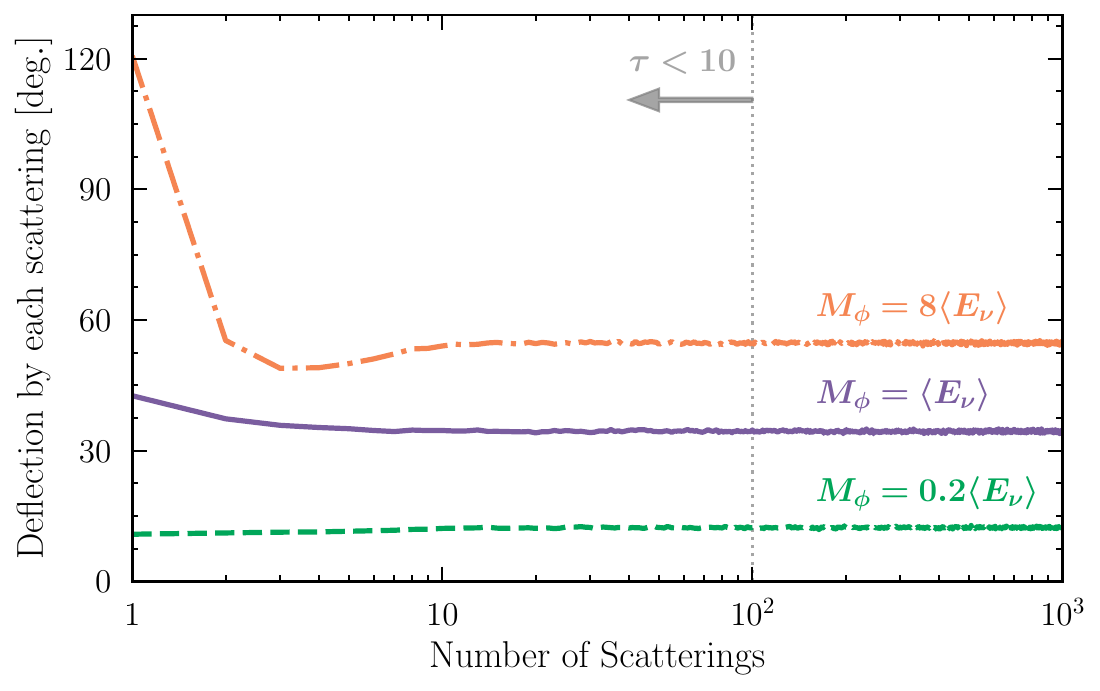}  
    \caption{Average {\it individual} deflection by each resonant scattering, for different $M_{\phi}/\langle E_\nu\rangle$.}
    \label{fig:def_each_scatt}
\end{figure}

Supplemental \Cref{fig:def_each_scatt} shows, for completeness, that the average deflection by each \emph{individual} scattering is sizable. It rapidly converges to a constant, as scattering processes tend to reach equilibrium. Also, deflections are bigger for larger $M_{\phi}/\langle E_{\nu} \rangle$ as discussed above.

To generate Suppl.~\cref{fig:def_total,fig:def_dist,fig:def_each_scatt}, we perform numerical simulations following these steps:
\begin{enumerate}
    \item We start with an ensemble of test neutrinos with energies $\{E_1\}$, randomly drawn from a Maxwell-Boltzmann distribution with average energy $\langle E_{\nu} \rangle$.
    \item Each test neutrino with energy $E_1$ scatters with a background neutrino with energy $E_2$ and direction $\cos \theta_{12}$. As the probability density for this scattering to happen is proportional to the optical depth in \cref{eq:tau_integral},
    \begin{equation}
        \mathcal{P}(E_2, \cos \theta_{12} | E_1) \propto  \,f(E_2,\hat{\Omega}_2) \,(1 - \cos \theta_{12}) \,\sigma_{\nu\nu}(E_1,E_2,\cos \theta_{12}) \, ,
    \end{equation}
    we draw $E_2$ and $\cos \theta_{12}$ from this distribution. Plugging the cross sections~(\ref{eq:csec_nres_LM}), (\ref{eq:csec_nres_SM}), and (\ref{eq:csec_res}); and setting $f(E_2, \hat{\Omega}_2)$ to an isotropic Maxwell-Boltzmann distribution with average energy $\langle E_\nu \rangle$, we have
    \begin{itemize}
        \item $M_\phi \ll \sqrt{s}$:
        \begin{equation}
        \mathcal{P}(E_2, \cos \theta_{12} | E_1) \propto  E_2 \, \exp\left[ - \frac{3 E_2}{\langle E_\nu \rangle}\right] \,;
        \label{eq:dist_SM}
        \end{equation} 
        \item $M_\phi \gg \sqrt{s}$:
        \begin{equation}
        \mathcal{P}(E_2, \cos \theta_{12} | E_1) \propto (1 - \cos\theta_{12})^2 \, E_2^3  \, \exp\left[ - \frac{3 E_2}{\langle E_\nu \rangle}\right]  \,;
        \label{eq:dist_LM}
        \end{equation}
        \item $M_\phi \sim \sqrt{s}$ (resonant regime):
        \begin{align}
        & \mathcal{P}(E_2, \cos \theta_{12} | E_1) \propto E_2^2 \, \exp \left[- \frac{3 E_2}{\langle E_\nu \rangle} \right]\,  \delta\left( E_2 -\frac{M_{\phi}^2}{2E_1 (1 - \cos\theta_{12})}\right) \nonumber
        \\
        & \qquad\qquad \propto  (1 - \cos\theta_{12})^{-2} ~ {\rm exp}\left[\frac{- 3 M_{\phi}^2}{2 E_1 \langle E_{\nu} \rangle (1 - \cos\theta_{12} )}  \right] \, \delta\left( E_2 -\frac{M_{\phi}^2}{2E_1 (1 - \cos\theta_{12})}\right) 
        \label{eq:dist_res}
        \end{align}
    \end{itemize}
    with $-1 \leq \cos \theta_{12} \leq 1$ and $E_2 > 0$.
    \item In each scattering, the angles between the test neutrino direction and the outgoing neutrino directions, $\Delta \theta$, are given by kinematics
    \begin{equation}
        \cos \Delta \theta = \frac{1}{\gamma_{\rm CM} (1 \pm \beta_{\rm CM} \,\cos\vartheta_c )} \left[ \pm~ {\rm sin}\,\theta_{1c} \,{\rm sin}\,\vartheta_c\, \cos\varphi_c + \gamma_{\rm CM}\, \cos\theta_{1c} (\beta_{\rm CM} \pm \cos\vartheta_c) \right] \, , 
        \label{eq:u_13_14}
    \end{equation}
    where $\cos\vartheta_c$ and $\varphi_c$ are the outgoing polar and azimuth angle in the CM frame, respectively; and $\theta_{1c}$ is the angle between the test neutrino momentum and the boost direction of the CM,
    \begin{equation}
        \cos\theta_{1c} = \frac{E_1/E_2 + \cos\theta_{12}}{\sqrt{ \left( E_1 /E_2 + \,\cos\theta_{12}\right)^2 + {\rm sin}^2\,\theta_{12}}}\,.
    \end{equation}
    We draw $\cos\vartheta_c$ and $\varphi_c$ from uniform distributions over the intervals $[-1,\,1]$ and $[0,\,2\pi)$, respectively, as s-channel scattering is isotropic in the CM. This is not the case for other scattering channels~\cite{Esteban:2021tub}. To be conservative, we set the individual deflection by each scattering to the smallest value among the two possibilities in \cref{eq:u_13_14}.
    \item We repeat steps $2$ and $3$ to simulate many scatterings. 
    
    The incoming test neutrino energies in the $n$-th scattering are set to the outgoing neutrino energies in the $(n-1)$-th scattering; that is, from kinematics, 
    \begin{equation}
        E_1^{\,(n)} \equiv \frac{E_1^{\,(n-1)}+E_2^{\,(n-1)}}{2} \, ( 1 \pm \beta_{\rm CM}^{\,(n-1)} \,\cos\vartheta_c^{\,(n-1)}) \, ,
    \end{equation}
    where the two values correspond to the two possible angles in \cref{eq:u_13_14}. To be consistent, we choose the energy corresponding to the smallest $\Delta \theta$.
    \item The \emph{total} deflection with respect to the initial direction after the $n$-th scattering is given by~\cite{Goudsmit:1940zza} 
    \begin{equation}
        \cos \theta_\mathrm{total}^{(n)} = \cos \theta_\mathrm{total}^{(n-1)} \cos \Delta \theta^{(n)} + \sin \theta_\mathrm{total}^{(n-1)} \sin \Delta \theta^{(n)} \cos \phi^{(n)} \, ,
        \label{eq:total_defl}
    \end{equation}
    with $\Delta \theta^{(n)}$ the individual deflection by the $n$-th scattering, \cref{eq:u_13_14}; and $\phi^{(n)}$ the relative azimuth deflection by the $n$-th scattering, uniformly distributed over $[0, \,2\pi)$.
    
    Repeatedly applying \cref{eq:total_defl}, we obtain the total deflection angle after many scatterings, $\theta_\mathrm{total}$.
\end{enumerate}
The results in Suppl.~\cref{fig:def_total,fig:def_dist,fig:def_each_scatt} are obtained with an ensemble of $10^4$ test neutrinos.

Realistically, decoupling happens when $\nu$SI no longer randomize neutrino directions. Using the simulation described above, we have computed the number of scatterings below which directions are no longer randomized, $N_\mathrm{scat}^\mathrm{rand}$ (as described above, when the number of scatterings is very large directions get readily randomized). We consider directions to be no longer randomized when the relative difference between the amount of forward-moving neutrinos ($\cos \theta_{\rm total}>0$) and backward-moving neutrinos ($\cos \theta_{\rm total}<0$) is larger than 10\%. The square root of $N_\mathrm{scat}^\mathrm{rand}$ gives $\tau_\mathrm{FS}$. Supplemental~\Cref{fig:bounds_FS_isotropic} shows that our choice in the main text, ${\tau_\mathrm{FS} = 10}$, is a good approximation to this.

\begin{figure}[hbtp]
    \centering
    \includegraphics[width=0.61\textwidth]{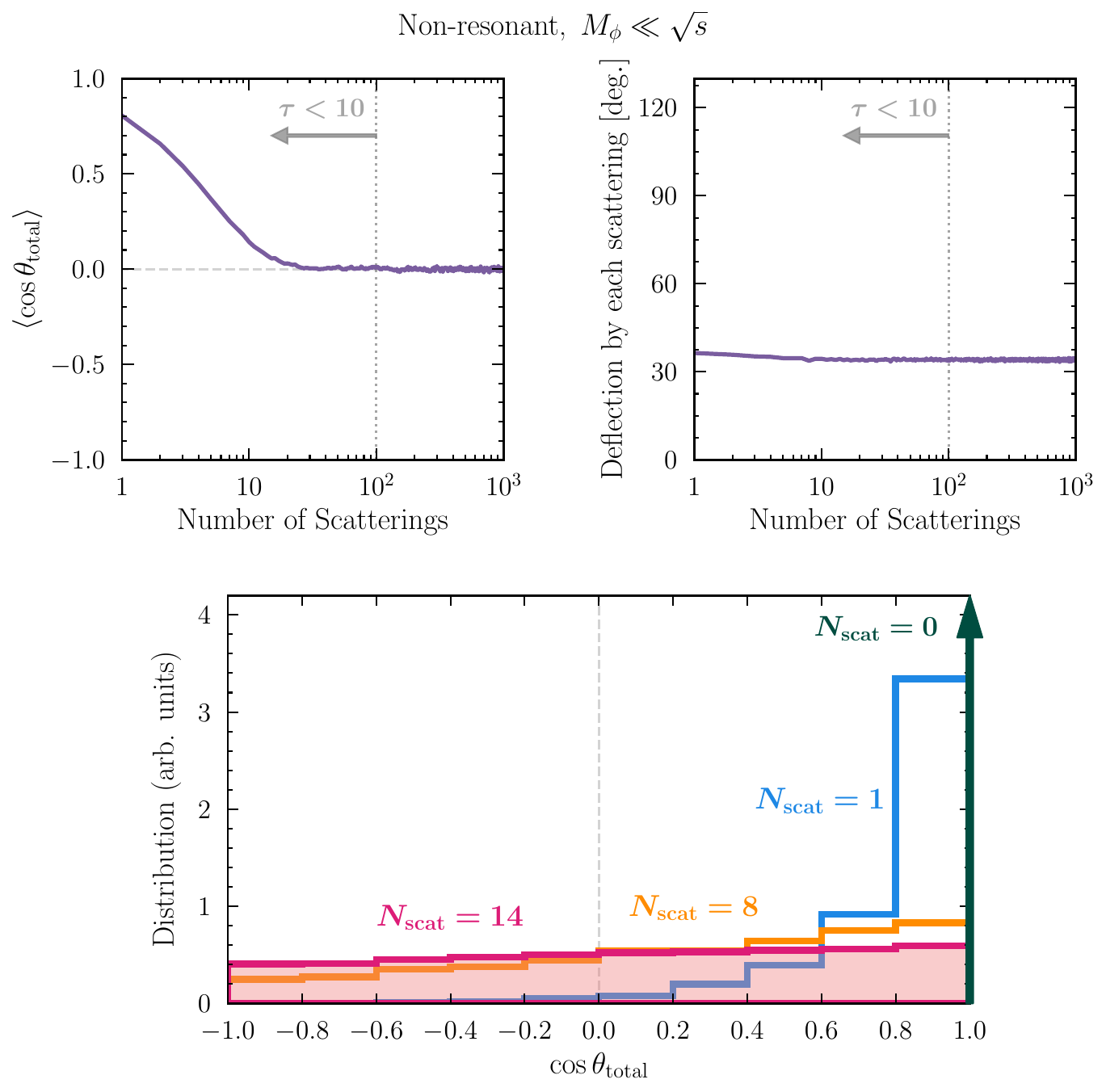}  
    \caption{Same as Suppl.~\cref{fig:def_total,fig:def_dist,fig:def_each_scatt}, but for non-resonant scattering with $M_{\phi} \ll \sqrt{s}$.}
    \label{fig:nres_SM_def}
\end{figure}

\begin{figure}[hbtp]
    \centering
    \includegraphics[width=0.61\textwidth]{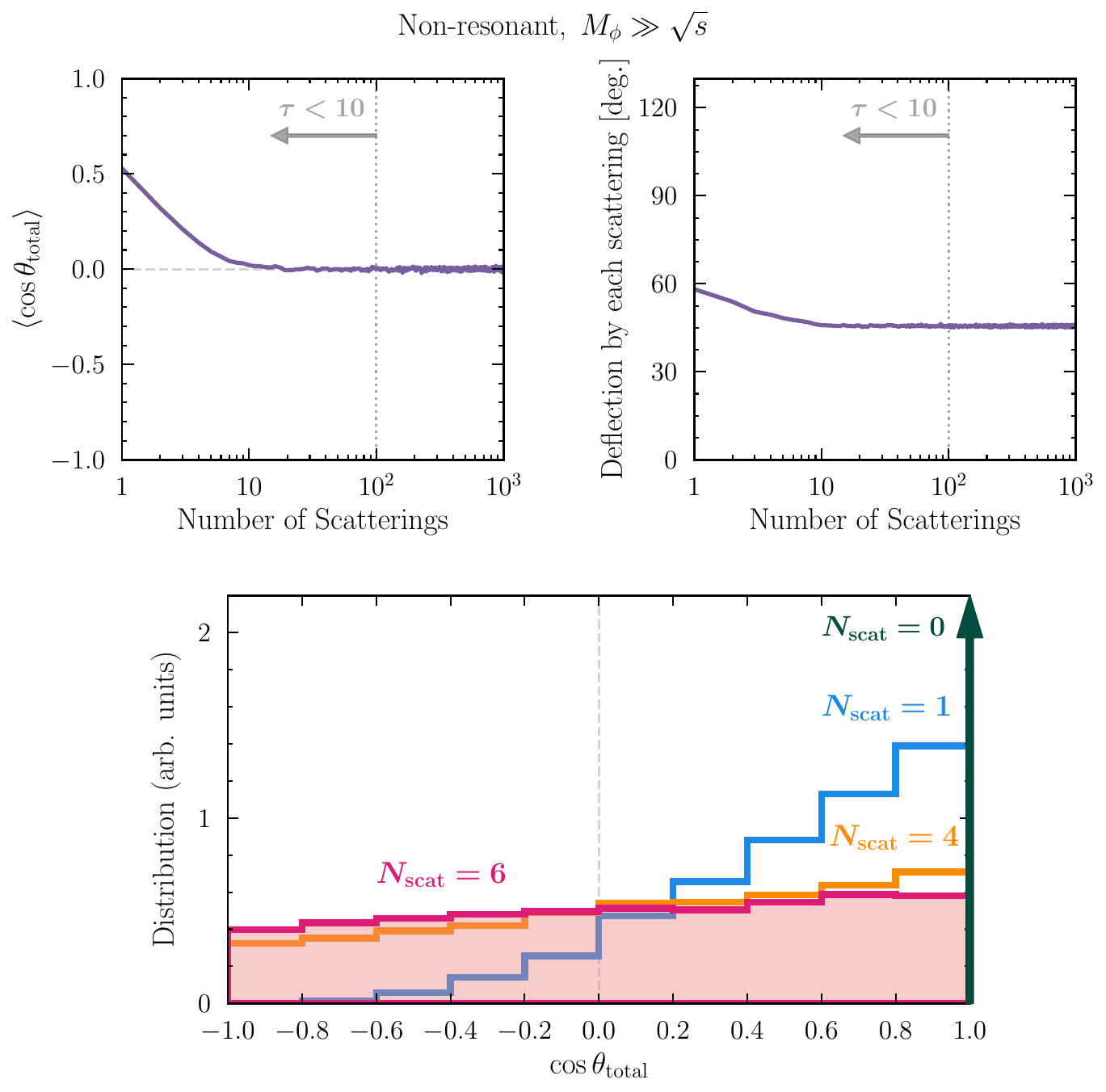}  
    \caption{Same as Suppl.~\cref{fig:def_total,fig:def_dist,fig:def_each_scatt}, but for non-resonant scattering with $M_{\phi} \gg \sqrt{s}$.}
    \label{fig:nres_LM_def}
\end{figure}

Supplemental~\Cref{fig:nres_SM_def,fig:nres_LM_def} show our results of our simulations for non-resonant scattering, which are insensitive to $M_{\phi}/\langle E_{\nu} \rangle$ (see~\cref{eq:dist_LM,eq:dist_SM}). 

\clearpage
\newpage

\section{Energy dependence of the decoupling condition}
\label{sec:supplemental_energy}

Here we assess the impact of the energy dependence of the $\nu$SI cross section, \cref{eq:csec}.  In the main text and the derivations in \cref{sec:supplemental_tau}, we obtain the size of the neutrino ball at decoupling, $\ell_\mathrm{FS}$, by requiring the \emph{average} optical depth of all neutrinos to be $\tau_\mathrm{FS}$ (see~\cref{eq:tau_integral}).

\begin{figure}[hbtp]
    \centering
    \includegraphics[width=0.9\textwidth]{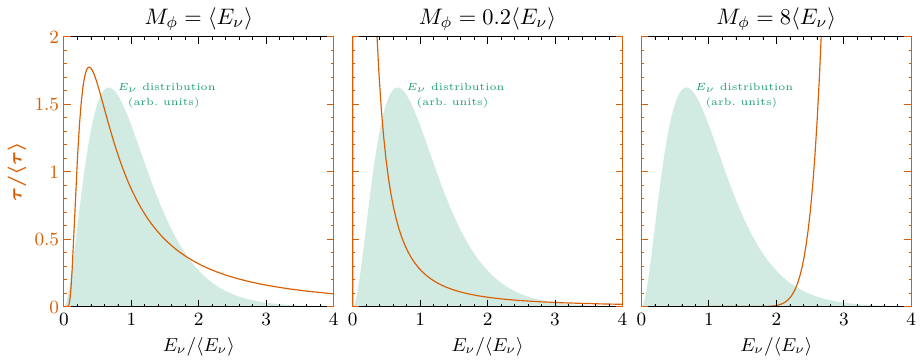}    
    \caption{Resonant optical depth as a function of neutrino energy, divided by its average value, for different mediator masses. For very small mediator masses, $\tau/\langle\tau\rangle$ is independent of $M_\phi$.}
    \label{fig:tau_energy}
\end{figure}

Supplemental~\Cref{fig:tau_energy} shows the fraction of neutrinos with optical depths $\tau$ close to the average $\langle \tau \rangle$. We compute $\tau$ assuming resonant scattering, averaging over the target neutrino distribution, which smooths out the strong energy dependence of the cross section ${\sigma_\mathrm{res} \propto \delta(s-M_\phi^2) = \delta(2 E_1 E_2 [1-\cos \theta_{12}] - M_\phi^2)}$. The figure also illustrates the rich phenomenology of energy-dependent $\nu$SI. If $M_\phi$ deviates from $\langle E_\nu \rangle$, part of the neutrinos will free-stream earlier ($\ell_\mathrm{FS} \propto \tau_{\rm FS}^{-2}$). In the burst-outflow case, different neutrino energies would have rather different signal durations.

These results do not affect our conservative SN 1987A sensitivity. Kam-II and IMB observed neutrinos with energies between $\sim 8 \, \mathrm{MeV}$ and $\sim 40 \, \mathrm{MeV}$~\cite{Kamiokande-II:1987idp, PhysRevD.38.448, Bionta:1987qt, IMB:1988suc}.  Hence, we conservatively require that low-energy \emph{and} high-energy neutrinos must have $\ell_\mathrm{FS}/c \lesssim \ell_0/c \sim 10 \, \mathrm{s}$. Further, if we increase $\tau_\mathrm{FS}$ and decrease $\ell_\mathrm{FS}^2$ by the same factor, the sensitivity stays the same (see~\cref{eq:gres}); we could as well define our conservative analysis by $\ell_\mathrm{FS}/c = 15 \, {\rm s}$ (still incompatible with data, see~\cref{fig:data}) and $\tau_\mathrm{FS} = 40$, guaranteeing a large optical depth when decoupling begins across different energies. 

\begin{figure}[hbtp]
    \centering
    \includegraphics[width=0.63\textwidth]{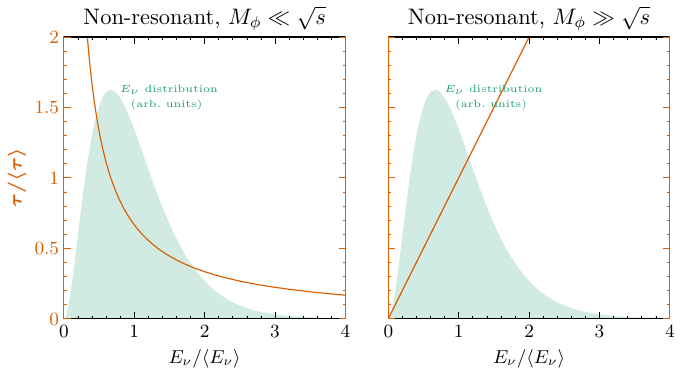}
    \caption{Same as Suppl.~\cref{fig:tau_energy}, but for non-resonant scattering.}
    \label{fig:tau_energy_off_resonant}
\end{figure}

Supplemental~\Cref{fig:tau_energy_off_resonant} shows that the picture is similar for non-resonant scattering. In this case the cross section has a weaker dependence on neutrino energy, hence the smoother optical depth.


\clearpage
\newpage

\section{Scenarios with Dirac neutrinos}
\label{sec:supplemental_Dirac}

Here we generalize our results to also cover Dirac neutrinos.  The simple $\nu$SI Lagrangian for Majorana neutrinos $\mathcal{L}_{\nu\mathrm{SI}} = -1/2 \, g \bar{\nu} \nu \phi$ admits two generalizations for Dirac neutrinos
\begin{align}
    \mathcal{L}^1_{\nu\mathrm{SI}, \, \mathrm{D}} & = - g \overline{\nu^c} \nu \phi + \mathrm{h.c.} \, ,
    \label{eq:nuSI_Lagrangian_Dirac_1} \\
    \mathcal{L}^2_{\nu\mathrm{SI}, \, \mathrm{D}} & =  - g \bar{\nu} \nu \phi \, ,
    \label{eq:nuSI_Lagrangian_Dirac_2}
\end{align}
with $\nu^c$ the charge conjugate of the neutrino field. In the first case, the mediator has lepton number $L_\phi=-2$~\cite{Burgess:1993xh}; whereas in the second case, $L_\phi=0$. 

The best sensitivity of supernova neutrinos to $\nu$SI comes from resonant scalar production. For the model in \cref{eq:nuSI_Lagrangian_Dirac_1}, lepton number conservation enforces this process to happen via $\nu \nu \rightarrow \phi^*$ or $\bar{\nu} \bar{\nu} \rightarrow \phi$. For the model in \cref{eq:nuSI_Lagrangian_Dirac_2}, it must happen via $\nu \bar{\nu} \rightarrow \phi$. As mentioned in the preamble of S.M., angular momentum conservation in the CM frame requires the initial particles to have the same helicity. Since supernova neutrinos are ultrarelativistic and are produced via Standard Model interactions, neutrinos are left-handed and antineutrinos right-handed. Thus, for the first model, resonant scattering takes place and our results in the resonant regime directly apply up to $\mathcal{O}(1)$ factors in the cross section~\cite{Esteban:2021tub}. 

For the second model, resonant scattering is only possible if a large population of right-handed neutrinos (and left-handed antineutrinos) builds up. This can happen via non-resonant t- and u-channel scattering, which flip the helicity of the interacting particles~\cite{Esteban:2021tub}. We estimate the frequency of these interactions by computing their associated mean free paths inside the PNS core. Neutrinos diffuse inside the core over timescales $\sim 1 \, \mathrm{s}$, as discussed in the main text. That is, they travel a total distance $d \sim c \cdot 1 \, \mathrm{s} \sim 3 \times 10^{10} \, \mathrm{cm}$. The neutrino number density is $n_\nu \sim 10^{36} \, \mathrm{cm}^{-3}$, corresponding to the thermal equilibrium density with an average energy $\langle E_\nu^\mathrm{core} \rangle \sim 100 \, \mathrm{MeV}$. Thus, the ratio between the traveled distance and the non-resonant $\nu$SI mean free path is
\begin{equation}
    \frac{d}{\lambda_\text{non-res}} = d \, n_\nu \, \sigma_\text{non-res} \sim 5 \times 10^4 \cdot \left( \frac{g}{5 \times 10^{-5}} \right)^4 \left( \frac{10 \, \mathrm{MeV}}{M_\phi} \right)^2 \, ,
\end{equation}
where we approximate the non-resonant scattering cross section as $\sigma_\text{non-res} \sim g^4 / (4 \pi M_\phi^2)$. This ratio is very large even for the smallest couplings of our conservative analysis (see~\cref{fig:bounds}), ensuring a large number of scatterings and thus that a large population of right-handed neutrinos and left-handed antineutrinos builds up. Thus, our conservative analysis holds.

For lower couplings, such as the lowest end of our estimated sensitivity ($g \sim 2 \times 10^{-6}$), $d / \lambda_\mathrm{non-res} \sim 0.1$. That is, a small but non-negligible population of right-handed neutrinos and left-handed antineutrinos will gradually build up. A more involved calculation taking into account all interaction channels (including, e.g., double-scalar production or scalar radiation in neutrino-nucleus interactions) and the time-dependent abundance of right-handed neutrinos is thus needed to understand the full $\nu$SI sensitivity of a dedicated analysis in this model. 


\clearpage
\newpage

\section{Relativistic hydrodynamics of a neutrino fluid}
\label{sec:supplemental_hydro}

Here we describe the relativistic hydrodynamic equations that govern the evolution of neutrinos with strong $\nu$SI, and we derive solutions for the \emph{burst outflow} and \emph{wind outflow} cases.

As discussed in the main text, in our region of interest the self-scattering mean free path of neutrinos is initially tiny, on the $\mu$m scale --- many orders of magnitude smaller than the size of the PNS and any other relevant length scale. Hence, the behavior of the neutrino ball is described by relativistic hydrodynamics of a perfect fluid.

Outside the PNS, the fluid equations follow from energy and momentum conservation~\cite{Weinberg:1972kfs},
\begin{equation}
    \nabla_{\alpha} T^{\alpha \beta} = 0  \,, \label{eq:EM_cons}
\end{equation}
with $T^{\alpha \beta}$ the energy-momentum tensor and $\nabla_\alpha$ the covariant derivative. The former can be related to the neutrino energy density $\tilde{\rho}$ and pressure $\tilde{P}$ in the comoving frame, i.e., the frame where the fluid is locally at rest,  
\begin{equation}
    T^{\alpha \beta} = \tilde{P}\, g^{\alpha\beta} + (\tilde{\rho} + \tilde{P}) \, U^{\alpha}U^{\beta} = \frac{1}{3}\tilde{\rho}\, g^{\alpha\beta} + \frac{4}{3}\tilde{\rho} \, U^{\alpha}U^{\beta}  \, , \label{eq:EM_tensor}   
\end{equation}
with $g^{\alpha \beta}$ the metric tensor, $U^\alpha$ the four-velocity of the fluid, and for a relativistic fluid $\tilde{P} = \tilde{\rho}/3$. $U^\alpha$ is related to the fluid bulk velocity $\vec{v}$ as
\begin{align}
    U^0 & = \gamma \, , \\
    U^{i} & = \gamma v^i \,,
\end{align}
with $\gamma = (1-|\vec{v}|^2)^{-1/2}$ the fluid bulk Lorentz factor.

As we only consider number-conserving $2 \rightarrow 2$ scattering, the total number of neutrinos is also conserved,
\begin{equation}
    \nabla_{\alpha} (\tilde{n} \, U^{\alpha}) =  0 \label{eq:n_cons_1}
\end{equation}
with $\tilde{n}$ the neutrino number density in the comoving frame, related to the laboratory-frame number density $n$ by $n = \gamma \tilde{n}$.

\Cref{eq:EM_cons,eq:n_cons_1} describe the flow of tightly coupled ultrarelativistic neutrinos in the absence of external forces. If we further assume spherical symmetry, the energy-momentum tensor components are given by
\begin{align}
    T^{00} & = \tilde{\rho}\gamma^2 \left(1 + \frac{1}{3}v^2 \right) \,,
    \\
    T^{0r} & = \frac{4}{3} \tilde{\rho}\gamma^2 v  \,,
    \\
    T^{rr} & = \tilde{\rho}\gamma^2 \left(\frac{1}{3} + v^2 \right) \,,
    \\
    T^{\vartheta \vartheta} & = \frac{1}{3} \tilde{\rho} r^{-2} \, , \\
    T^{\varphi\varphi} & = \frac{1}{3} \tilde{\rho} r^{-2}  \sin^{-2} \vartheta \,.
\end{align}
Inserting the proper Christoffel symbols to compute the covariant derivatives in spherical coordinates, \cref{eq:EM_cons,eq:n_cons_1} read
\begin{align}
    & \frac{\partial}{\partial t} \left[ \tilde{\rho}\gamma^2 \left( 1 + \frac{1}{3}v^2\right) \right] + \frac{1}{r^2} \frac{\partial}{\partial r} \left( r^2 \frac{4}{3}\, \tilde{\rho}\gamma^2 v  \right) = 0 \,, \label{eq:E_cons}
    \\
    & \frac{\partial}{\partial t} \left( \frac{4}{3} \,\tilde{\rho}\gamma^2 v \right) + \frac{1}{r^2}  \frac{\partial}{\partial r} \left[ r^2 \tilde{\rho}\gamma^2 \left(\frac{1}{3} + v^2 \right)  \right] - \frac{2}{3r} \tilde{\rho} = 0 \,, \label{eq:M_cons}
    \\
    & \frac{\partial n}{\partial t} + \frac{1}{r^2}\frac{\partial}{\partial r} \left(r^2 n v \right) = 0  \,. \label{eq:n_cons_3}
\end{align}
\Cref{eq:E_cons,eq:M_cons,eq:n_cons_3} are our starting point to derive the different supernova neutrino outflows with strong $\nu$SI.

The \emph{burst outflow} happens when a uniform neutrino fluid undergoes free expansion in vacuum. Given uniform initial conditions for a neutrino ball of size $\ell_0$ whose edge is expanding at the speed of light, i.e., $n(r, t=0) = n_0 H(\ell_0 - r)$, $\partial_t n(r, t=0) = \dot{n}_0 H(\ell_0 - r)$, and $v(r=\ell_0, t=0) = 1$, with $H$ the Heaviside step function; we obtain the following solution to \cref{eq:E_cons,eq:M_cons,eq:n_cons_3}
\begin{align}
    n(r, t) & = n_0 \left( \frac{\ell_0}{\ell(t)} \right)^3 H(\ell(t) - r) \, , \label{eq:homologous_n}\\
    v(r, t) & = \frac{r}{\ell(t)} \, , \label{eq:homologous_v}
\end{align}
with $\ell(t) = \ell_0 + t$. This is the homologous expansion of a homogeneous neutrino ball, as described in the main text.

We have checked with the \texttt{PLUTO} hydrodynamics code~\cite{Mignone_2007} that a variety of smooth initial density and velocity profiles evolve after a short time as \cref{eq:homologous_n,eq:homologous_v}. This is consistent with the behavior found in similar studies of free expansions of relativistic gases~\cite{1980Ap&SS..72..447Y}. Intuitively, the form of the velocity profile can be understood by integrating \cref{eq:n_cons_3} from $r$ to $\ell$,
\begin{equation}
     \ell^2 \, n(\ell,\,t) \, v(\ell,\,t) - r^2\, n(r,\,t) \,v(r,\,t) =  - \int_r^{\ell} {\rm d}r' \, r'^2 \,\frac{\partial n}{\partial t}  \,. \label{eq:n_cons_4}
\end{equation}
This equation is the integral form of the continuity equation, reflecting number conservation in a spherical shell. If both $n$ and $\partial n/\partial t$ are approximately uniform and different from 0, we obtain $v(r) \propto r$. Finally, we note that \cref{eq:homologous_n} has a sharp discontinuity at $r=\ell$, where \cref{eq:homologous_v} indicates an expansion velocity exactly equal to $c$. Hence, close to the edge the solution must change to a smooth transition to $n=0$ and $v \neq c$. This can also be seen from the laboratory-frame energy density profile associated to the homologous expansion, ${T^{00}(r, t) \propto \frac{3 + (r/\ell(t))^2}{[1 - (r/\ell(t))^2]^3}}$, whose integral over the ball diverges. In our \texttt{PLUTO} simulations we observe a smooth transition to vacuum outside the ball, a result also found in Ref.~\cite{1980Ap&SS..72..447Y}. The details of this transition do not affect our results, as we aim for factor-two precision and we conservatively compute the optical depth for a neutrino traveling a distance $\ell$, hence our results are not sensitive to how the edge of the ball expands.

Alternatively, the {\it wind outflow} of a neutrino fluid occurs when we look for a steady-state solution to the equations of hydrodynamics. Such solutions have been invoked in fireball models of gamma-ray bursts~\cite{Piran:1993jm} and in Parker's solar-wind model~\cite{1965SSRv....4..666P}. If we drop the time-derivative terms in \cref{eq:E_cons,eq:M_cons,eq:n_cons_3}, we obtain
\begin{align}
    \frac{\partial}{\partial r} \left( r^2 \frac{4}{3}\, \tilde{\rho}\gamma^2 v  \right) & = 0 \,, \label{eq:E_cons_wind}
    \\
    \frac{1}{r^2}  \frac{\partial}{\partial r} \left[ r^2 \tilde{\rho}\gamma^2 \left(\frac{1}{3} + v^2 \right)  \right] - \frac{2}{3r} \tilde{\rho} & = 0 \,, \label{eq:M_cons_wind}
    \\
    \frac{\partial}{\partial r} \left(r^2 n v \right) & = 0  \,. \label{eq:n_cons_wind}
\end{align}
Combining \cref{eq:E_cons_wind,eq:M_cons_wind} gives a first-order ordinary differential equation for $v(r)$
\begin{equation}
\frac{1}{4}\frac{{\rm d} v}{{\rm d} r} \left( \frac{1}{v^2} - 3 \right) + \frac{1}{2r\gamma^2 v} = 0 \, , \label{eq:EM_cons_steady}
\end{equation}
whose solution is
\begin{equation}
    r \sqrt{v (1-v^2)} = {\rm constant} \,. \label{eq:sonic}
\end{equation}
The function $\sqrt{v (1-v^2)}$ is concave with a maximum at $v=1/\sqrt{3}$. This function must decrease as $r$ increases, so depending on the boundary conditions there are two possible solutions. If $v < 1/\sqrt{3}$, the fluid decelerates as $r$ increases and, at large $r$, $v(r) \propto r^{-2}$. If $v > 1/\sqrt{3}$ the fluid accelerates as $r$ increases and, at large $r$, $\gamma(r) \propto r$ --- i.e., $1-v(r) \propto 1/r^2$. From \cref{eq:n_cons_wind,eq:E_cons_wind}, the former implies constant density and pressure at $r \rightarrow \infty$ and it is hence unphysical in the absence of external pressure, whereas the latter implies $n(r) \propto r^{-2}$ at large $r$ and no pressure at $r \rightarrow \infty$. This corresponds to the wind outflow described in the main text. In the wind outflow, the bulk velocity quickly approaches $c$, individual neutrinos move radially, and the arguments that make $\nu$SI burst outflows increase the duration of the supernova neutrino signal do not apply, as described in the main text.

\begin{figure}[hbtp]
    \centering
    \includegraphics[width=0.75\textwidth]{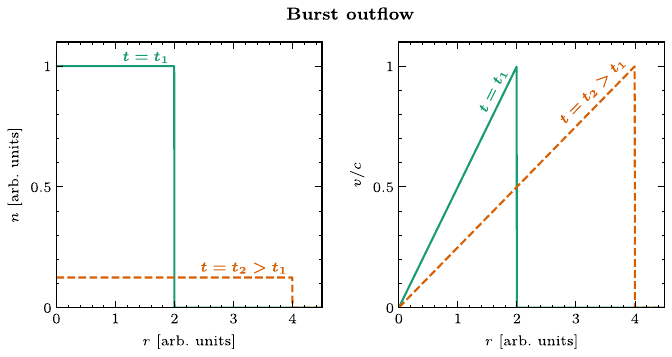}
    \includegraphics[width=0.75\textwidth]{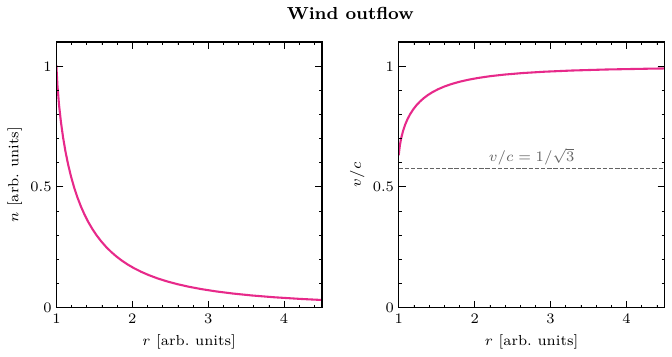}
    \caption{Laboratory-frame density $n(r)$ and expansion velocity $v$ profiles for the burst-outflow and wind-outflow cases. The wind outflow is time-independent. \emph{Unlike burst outflows, the wind outflow is inhomogeneous and expands with bulk velocities very close to $c$}.}
    \label{fig:outflows}
\end{figure}

Supplemental~\Cref{fig:outflows} shows the density and velocity profiles for both outflow cases. The wind outflow is the only physical steady-state outflow, requires $v > 1/\sqrt{3}$ everywhere, and cannot be continued down to $r=0$ where \cref{eq:sonic} has no solution. Hence, it requires a boundary condition at finite $r$ with $v > 1/\sqrt{3}$. For supernova neutrinos, this condition must be set at the PNS, where neutrino production and scattering with baryons modify \Cref{eq:EM_cons,eq:n_cons_1}. To gain insight, below we investigate the outflow close to the edge of the PNS in a simplified model. More detailed investigation, beyond the scope of this work, is needed to fully understand the problem.

As a first approximation to the physics close to the edge of the PNS, we allow for momentum transfer between the neutrino fluid and the baryons. We model this by introducing a bulk neutrino momentum loss with a rate inversely proportional to the time between neutrino-nucleon collisions, i.e., the neutrino-nucleon mean free path $\lambda_{\nu N}$. We set the net energy transfer between the neutrinos and baryons to zero, in keeping with the steady state assumption. (The technical assumption is that the net neutrino heating or cooling of the baryons in the layer of interest is negligible compared to the neutrino luminosity. This assumption will be violated in the deep interior of the PNS, but it is a reasonable first approximation in the outer layers since they have a small fraction of the overall heat capacity.) This modifies \cref{eq:EM_cons} as
\begin{align}
    \nabla_{\alpha} T^{\alpha t} &= 0 \, , \label{eq:drag_1}\\
    \frac{1}{T^{0r}}\nabla_{\alpha} T^{\alpha r} & = - \frac{1}{\lambda_{\nu N}(r)} \, . \label{eq:drag_2}
\end{align}
If we look for a steady-state solution to \cref{eq:drag_1,eq:drag_2}, we obtain, following the same steps as above,
\begin{equation}
\frac{1}{4}\frac{{\rm d} v}{{\rm d} r} \left( \frac{1}{v^2} - 3 \right) + \frac{1}{2 r \gamma^2 v} = \frac{1}{\lambda_{\nu N}(r)} \, . \label{eq:drag_3}
\end{equation}

Outside the PNS, $1/\lambda_{\nu N}=0$, recovering \cref{eq:EM_cons_steady} and, if $v > 1/\sqrt{3}$, the wind outflow. Inside the PNS, $\lambda_{\nu N} \ll r$, and \cref{eq:drag_3} gives $v \rightarrow 1/(4\tau_{\nu \mathrm{N}})$, where $\tau_{\nu \mathrm{N}} = (R-r)/\lambda_{\nu N}$ is the neutrino-nucleon optical depth into the PNS. As we go down into the PNS, within a few optical depths $v$ becomes $\ll 1$. This solution also has an energy density that increases linearly with $\tau$, which is similar to the case of radiative diffusion \cite{1960ratr.book.....C}. As we go outward, in order to match with the wind outflow, $v=1/\sqrt{3}$ must be crossed continuously. This can only happen if $2 r \gamma^2 v = \lambda_{\nu N}$, i.e., for $r=\lambda_{\nu N}(r)/\sqrt{3}$, at the edge of the PNS where the neutrino mean free path is of the order of the size of the PNS.

\begin{figure}[hbtp]
    \centering
    \includegraphics[width=0.6\textwidth]{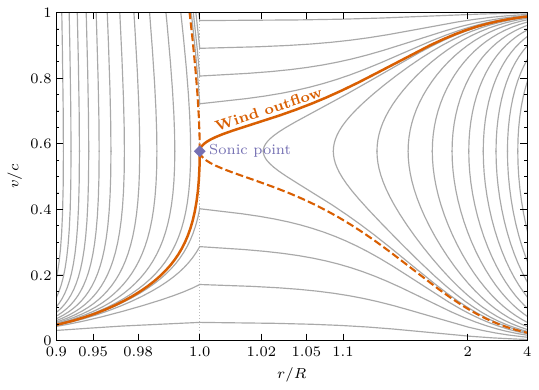}
    \caption{Steady-state outflows \emph{outside and inside} the PNS. The $x$-axis is arcsinh-stretched to show details near $r\approx R$. \emph{The wind outflow is the only steady-state solution inside and outside the PNS, and it must have $v=1/\sqrt{3}$ at the edge}.}
    \label{fig:wind}
\end{figure}

Supplemental~\Cref{fig:wind} shows the solutions to \cref{eq:drag_3}, assuming $\lambda_{\nu N} = R/50$ inside the PNS and $1/\lambda_{\nu N}=0$ outside (the behavior is qualitatively similar for other values of $1/\lambda_{\nu N} \gg 1/R$), where $R$ is the PNS radius. The sonic point at the edge of the PNS corresponds to the only point where the solution can be single-valued, continuous, and cross $v=1/\sqrt{3}$. The solid orange line is the only continuous steady-state outflow with no pressure at $r \rightarrow \infty$, it therefore corresponds to the \emph{wind outflow}. In it, the fluid must rapidly accelerate at the edge of the PNS, from $v \simeq 0.14$ to $v=1/\sqrt{3} \simeq 0.58$ in the last neutrino-nucleon mean free path (from $r/R=0.98$ to $1.0$). Acceleration outside the PNS is more gradual.

These results imply that the wind outflow requires unique conditions \emph{inside and outside the PNS}. Further work is needed to understand if generic initial conditions relax to the wind solution, the associated timescales, and the impact on supernova physics. In the main text, we outline several ideas that may lead to new observables.

\end{document}